\documentclass[reprint]{nature-with-graphics} 

\usepackage{amsmath}
\usepackage{enumerate,bm}
\usepackage{graphicx}
\usepackage{grffile} 
\usepackage{color}
\usepackage{esint}
\usepackage{textpos}
\usepackage[usenames,dvipsnames]{xcolor}
\usepackage{subfigure}
\usepackage{flushend}
\usepackage{tabularx}
\usepackage{amssymb}
\usepackage{float}
\usepackage{subfigure}
\usepackage{upgreek}
\usepackage{hyperref}

\spacing{1.4} 

\title{Correlation-driven sub-3~fs charge migration in ionised adenine}

\author{Erik P. M{\aa}nsson$^{1,2}$,
Simone Latini$^3$,
Fabio Covito$^3$,
Vincent Wanie$^{1,2,4}$,
Mara Galli$^{1,5}$,
Enrico Perfetto$^{6,7}$,
Gianluca Stefanucci$^{7,8}$,
Hannes H\"ubener$^3$,
Umberto De Giovannini$^{3,9}$
Mattea C. Castrovilli$^{2,10}$,
Andrea Trabattoni$^1$,
Fabio Frassetto$^{11}$,
Luca Poletto$^{11}$,
Jason B. Greenwood$^{12}$,
Fran\c{c}ois L{\'e}gar{\'e}$^{4}$,
Mauro Nisoli$^{2,5}$,
Angel Rubio$^{3,13}$ 
and Francesca Calegari$^{1,2,14\ast}$
}

\begin{document}
\maketitle

\begin{affiliations}
\item ~Center for Free-Electron Laser Science, DESY, Notkestr.~85, 22607 Hamburg, Germany.
\item ~Inst.~for Photonics and Nanotechnologies CNR-IFN, P.za L.\,da\,Vinci 32, 20133 Milano, Italy.
\item ~Max Planck Institute for the Structure and Dynamics of Matter and Center for Free Electron Laser Science, 22761 Hamburg, Germany.
\item ~INRS-EMT, 1650 Blvd. Lionel Boulet J3X 1S2, Varennes, Canada.
\item ~Department of Physics, Politecnico di Milano, Piazza L.\,da\,Vinci 32, 20133 Milano, Italy.
\item ~CNR-ISM, Division of Ultrafast Processes in Materials (FLASHit), Area della ricerca di Roma~1, Via Salaria Km 29.3, I-00016 Monterotondo Scalo, Italy.
\item ~Dipartimento di Fisica, Universit{\`a} di Roma Tor Vergata, Via della Ricerca Scientifica, 00133 Rome, Italy.
\item ~INFN, Sezione di Roma Tor Vergata, Via della Ricerca Scientifica 1, 00133 Roma, Italy
\item ~Dipartimento di Fisica e Chimica, Universit{\`a} degli Studi di Palermo, Via Archirafi 36, I-90123, Palermo, Italy.
\item ~Inst.~for the Structure of Matter CNR-ISM, Area Ricerca di Roma1, Monterotondo, Italy.
\item ~Inst.~for Photonics and Nanotechnologies CNR-IFN, Via Trasea 7, 35131 Padova, Italy.
\item ~Centre for Plasma Physics, School of Maths and Physics, Queen's University Belfast, BT7 1NN, UK.
\item ~Center for Computational Quantum Physics (CCQ), The Flatiron Institute, 162~Fifth avenue, New York NY 10010, USA.
\item ~Institut fur Experimentalphysik, Universit{\"a}t Hamburg, Luruper Chaussee 149, D-22761 Hamburg, Germany.
\item[*] E-mail: francesca.calegari@desy.de.
\end{affiliations}

\spacing{1.7} 

\begin{abstract} 
Sudden ionisation of a relatively large molecule can initiate a correlation-driven process dubbed charge migration\cite{Cederbaum1999}, where the electron density distribution is expected to rapidly change. Capturing this few-femtosecond/attosecond charge redistribution represents the real-time observation of the electron correlation in the molecule. So far, there has been no experimental evidence of this process.  
Here we report on a time-resolved study of the correlation-driven charge migration process occurring in the bio-relevant molecule adenine after ionisation by a 15--35~eV attosecond pulse\cite{Calegari_2015}. We find that, the production of intact doubly charged adenine -- via a shortly-delayed laser-induced second ionisation event -- represents the signature of a charge inflation mechanism resulting from the many-body excitation. This conclusion is supported by first-principles time-dependent simulations. Our findings opens new important perspectives for the control of the molecular reactivity at the electronic time scale. 
\end{abstract}

The interaction of ionising radiation with molecules often leads to an internal electronic rearrangement, governed by correlated processes such as shake-up or Auger\cite{Drescher2002,Hanna2017} and Interatomic Coulombic Decay (ICD)\cite{Averbukh2011}. The superposition of electronic states resulting from the many-body excitation has been predicted to initiate attosecond charge migration along the molecular backbone, when the nuclei can be considered as frozen \cite{Cederbaum1999, Kuleff2013, Kuleff2014, Remacle2006}. In the last few decades, advances in the development of extreme ultraviolet (XUV) attosecond light sources \cite{Nisoli_2017, Calegari_2015} have given access to electron migration in molecules, holding a great promise for attochemistry. In this context, a signature of few-femtosecond charge dynamics has been identified in aromatic amino acids by exploiting the characteristic broadband of the attosecond radiation to create a coherent superposition of cationic eigenstates \cite{Calegari2014, Lara-Astiaso2018}. Nevertheless, the originally conceived charge migration process, depicted as a non-stationary charge distribution resulting from the removal of an electron from a correlated state, remains to be demonstrated. This would constitute a unique route for mapping in real-time the energy flow occurring from the single excited electron to all other coupled electrons in the molecule, i.e. the electron correlation \cite{Stolow2013}. The above-described ``purely electronic'' scenario would only survive until the nuclei start to move, i.e. typically less than 10~fs\cite{Jenkins2016, Polyak2018, Lara-Astiaso2016}. Therefore, one would need to act on the system very promptly after the ionisation event to take advantage of this fast charge redistribution and achieve control over the molecular reactivity. On a time scale of several tens of femtoseconds multi-electronic and non-adiabatic effects are fully entangled and their interplay has been recently identified in the relaxation dynamics following XUV-induced ionisation of organic molecules \cite{Marciniak2019,Herve2020}. 

In this work, we report on the first experimental evidence of correlation-driven charge migration, occurring in the nucleic-acid base adenine after sudden ionization by an XUV attosecond pulse. Our few-femtosecond time-resolved study has the motivation of tracking many-body effects in real-time \emph{before} non-adiabatic effects take place and potentially take advantage of them to obtain an ultrafast control "knob" for the molecular dissociation.

\begin{figure}[b]
\centering
\includegraphics[width=1\textwidth]{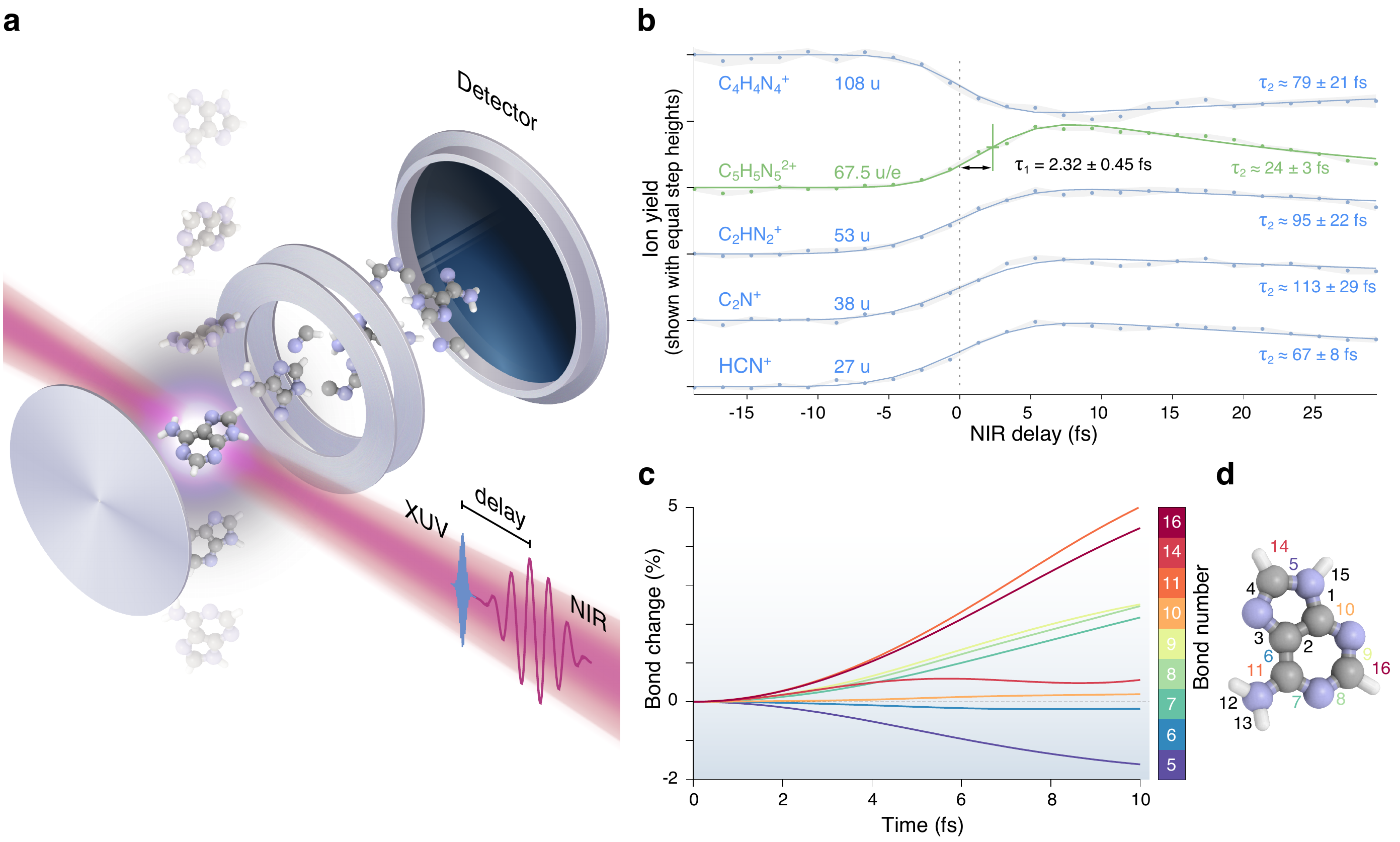}
\caption{\label{fig1}
\textbf{Experimental results: pump--probe scan.}
(a) Schematic of the experiment: a molecular beam is injected into a VMI operated in the ion time-of-flight mode. Adenine ions are accelerated towards the detector and the ion yield is measured as a function of the XUV-pump IR-probe delay.  
(b) Normalised yield of several ions (shown with a vertical offset) as a function of the XUV-pump NIR-probe delay. The yield of the ionic fragments exhibits a distinct positive or negative step-like behaviour, while the adenine parent dication (67.5 u/e) is fitted to have an exponential risetime of $\tau_1=2.32\pm0.45$\,fs (68\,\% confidence interval).
The decay lifetime ($\tau_2$) is significantly shorter for the dication (green curve) than for the cations. The grey shading indicates the standard error of the mean of 7 successive scans.
(c) Example of calculated time evolution of bond lengths in the first 10fs following XUV ionisation, with an electron removed from the fourth highest occupied molecular orbital (HOMO$-3$). All the bonds start elongating only after 4~fs. The theoretical simulations for bond elongation are performed with TDDFT.
(d) The bond numbering used in the theoretical work.
}\end{figure}

In our experiment, ionisation of adenine is initiated by an isolated sub-300~as XUV pulse containing photon energies from 15 to 35~eV, produced by high-harmonic generation \cite{Corkum1994} in krypton gas through the polarisation gating technique\cite{Sola2006}. A waveform-controlled 4-fs near-infrared (NIR, central photon energy 1.77~eV) probing pulse is combined with the XUV pump pulse using an interferometric approach. Adenine is sublimated and carried to the laser interaction region by using helium as a buffer gas. The produced ions are then collected as function of the XUV-pump NIR-probe delay (see Figure~\ref{fig1}a), using a time-of-flight spectrometer.
The ion mass spectrum resulting from ionisation by the XUV pulse is dominated by ionic fragments (81\% of the total yield as calculated in the Supplementary Information (SI)), indicating a relatively low photostability of the molecule in this energy range. Further deposition of energy by the NIR pulse, simultaneously or after the XUV, leads to an overall increase of fragmentation\cite{Pilling2007}. Figure~\ref{fig1}b shows the partial ion yield for several ionic fragments as a function of the pump-probe delay. The time dependent yields of the cationic fragments with mass 27, 38 and 53~u display step-like increases, followed by slower decay. The enhancement of small fragment ions occurs at the expense of the large fragments, mainly 108~u, which clearly indicates that the combination of XUV and NIR pulses leads to further excitation and therefore more efficient bond breaking.


\begin{figure}
\centering
\includegraphics[width=0.7\textwidth]{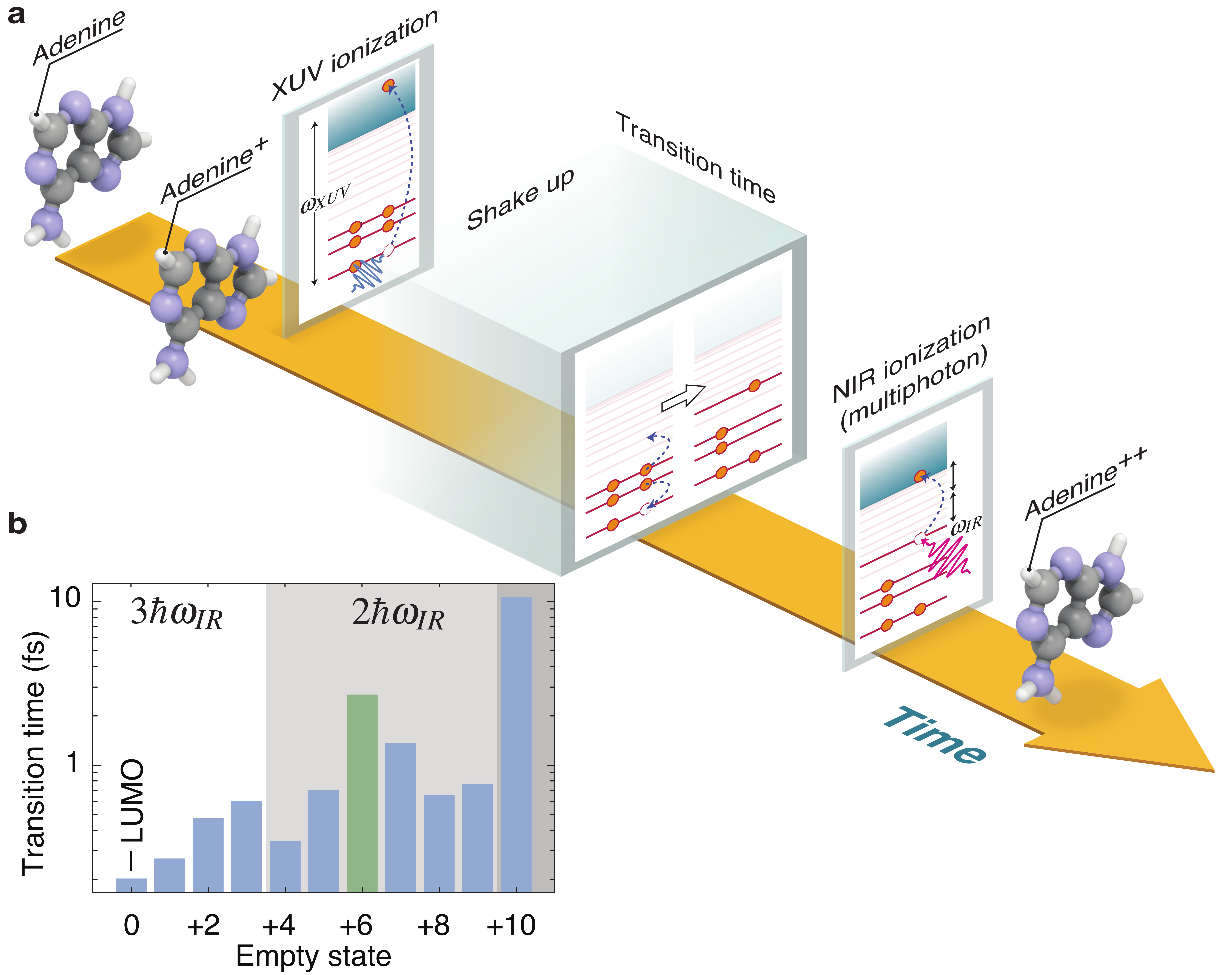} 
\caption{\label{fig2}
\textbf{Overview of the molecular dynamics: the shake-up process.}
(a) Following XUV photoionisation, a hole is created in the inner valence. The hole decays in a characteristic "transition time" and, due to electronic correlations, this can lead to excitation of a second electron to a bound excited state, called "shake-up" state. If optimally time-delayed from the XUV, the NIR control pulse extracts the excited electron, hence doubly ionising the molecule. (b) Transition times to a given shake-up state calculated with a Fermi's Golden rule approach. A special shake-up state (LUMO+6) is highlighted in green and shows a characteristic time of 2.5~fs. The states are ordered by energy and grouped (shades of gray) by the number of NIR photons (1, 2 or 3) required to ionise a second electron. The LUMO+6 state falls in the two NIR-photon group.}\end{figure}

The most intriguing observation in the time-dependent mass spectrum is the appearance of a new ion for small positive delays at mass/charge = 67.5~u/e, corresponding to the doubly charged parent molecule (adenine$^{2+}$)\cite{Alvarado2007, Bredy2009, Burgt2015}. It is worth noting that a stable dication of the parent is difficult to discern in the XUV-only signal or at negative NIR delays, and none is present if we select the portion of the XUV spectrum below 17\,eV (see the SI). Figure~\ref{fig1}b shows that the formation of the parent dication is delayed compared to the cationic fragments. Fitting the experimental data using a curve model described in the SI, we obtain for the dication pump--probe signal a delay of $2.32\pm0.45$\,fs ($\tau_1$, exponential risetime) and a decay time of $24 \pm 3$\,fs ($\tau_2$, exponential decay). To further verify that the steps of the cationic fragments accurately represent the absolute zero time delay (XUV--NIR overlap), we also did measurements with simultaneous injection of an atomic gas (krypton). The XUV+NIR double ionisation of krypton gives a Kr$^{2+}$ signal at a time consistent with the adenine cations (see SI), allowing us to conclude that it is the adenine dication signal which is positively delayed. The extracted delay does not appear to depend on the NIR pulse intensity, within the explored range from 7$\times 10^{12}$ to 1.4$\times 10^{13}$~W/cm$^2$. At the same time we observed that the dication yield scales quadratically with the NIR intensity (see SI), thus indicating that two NIR photons are required to doubly ionise the molecule.

The detection of a doubly charged ion with a sub-3~fs delay is a probe for a \emph{pure electronic mechanism} initiated by the XUV sudden ionisation and occurring \emph{before} the control NIR laser pulse arrives. It is worth mentioning that, the possibility of combined electronic and nuclear dynamics (non-adiabatic effects such as conical intersections \cite{Galbraith2017}) cannot be ruled out a priori. Nevertheless, in the sub-3~fs time window, we do not expect these effects to be significant.
This conclusion is supported by first-principles calculations\cite{Marques2003, Castro2006, Andrade2015} -- based on time-dependent density functional theory (TDDFT)\cite{runge_density-functional_1984,bertsch_real-space_2000} and Ehrenfest dynamics\cite{Alonso:2008fz} -- indicating that the nuclei can be almost considered as frozen in this short time scale. As an example, in Figure~\ref{fig1}c we report the time evolution of the bond lengths after removal of an electron from the fourth highest occupied molecular orbital (a similar analysis where the electron is removed from different occupied orbitals is reported in the SI and leads to the same conclusions). Several bond lengths are seen to evolve and drift away from their equilibrium value only after 4~fs.


\begin{figure}[bt!]
\centering
\includegraphics[width=0.7\textwidth]{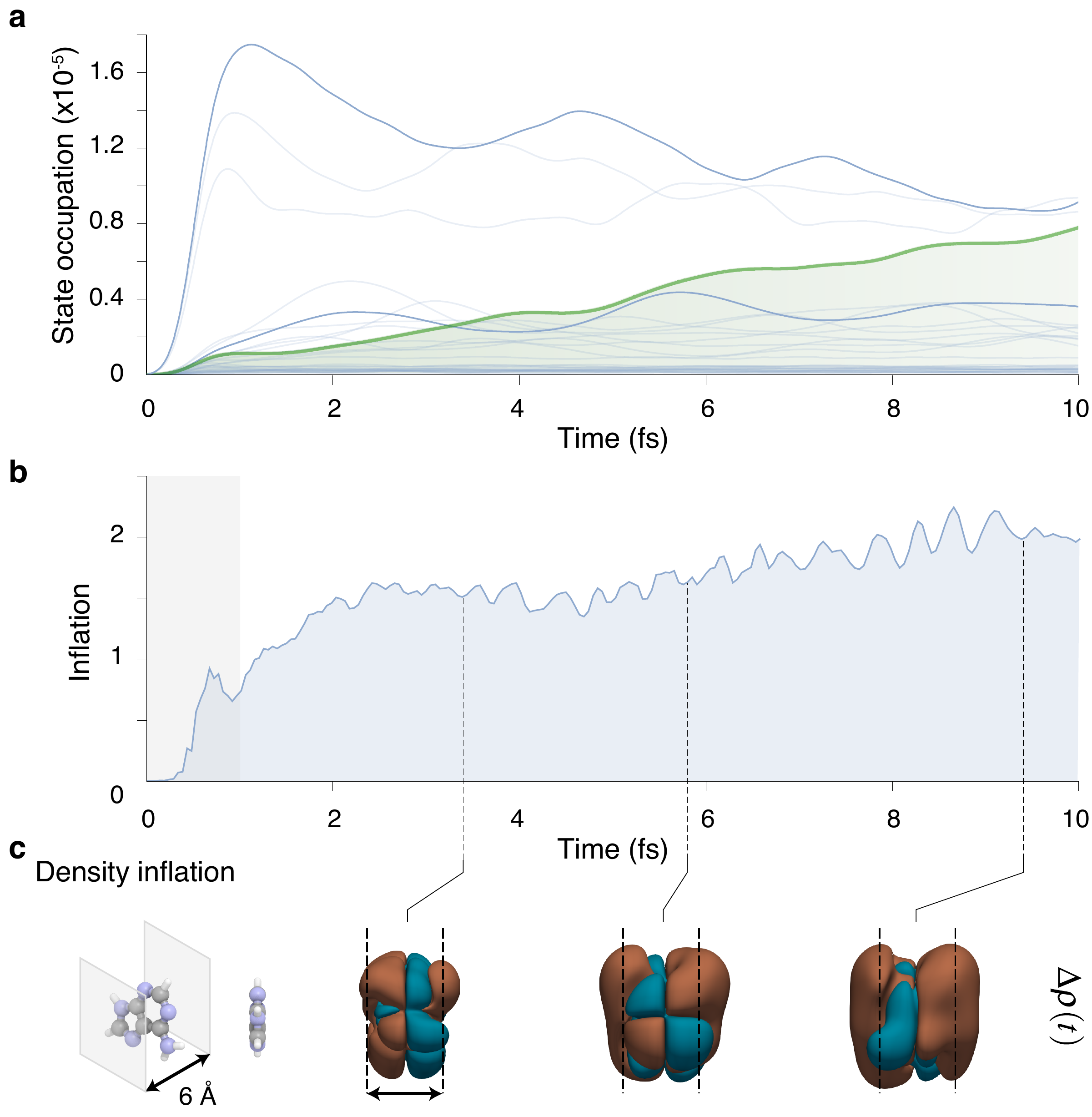}
\caption{\label{fig3}
\textbf{Theoretical results: shake-up and charge inflation.} (a) Time-dependent occupations of the adenine bound excited states after photoionisation by the XUV pulse, calculated with the ab-initio non-equilibrium Green's function method. The special state (LUMO+6), highlighted in green, is populated via the shake-up process and its population rises over several femtoseconds to one of the largest values. (b) Integrated time-dependent electron density more than 3~{\AA} away from the molecular plane. The grey shaded area represents the time-window of the pump pulse, having its peak at $t_\mathrm{pump}=0.48$\;fs. The y-axis has been rescaled by a factor of $10^5$.
(c) Left panel: the adenine molecule and the planes defining the integration region. Right panel: snapshots of the variation of the electronic density with respect to the density immediately after the XUV pulse. We observe that the electronic density inflates considerably away from the molecular plane. The noticeable out-of-plane charge migration can be attributed to the increasing population of the correlated LUMO+6 state (see SI).}
\end{figure}


Having in mind that the correlated electron dynamics may have a primary role in the delayed creation of the dication, we propose the following mechanism: \textit{I)} the XUV pulse singly ionises the molecule leaving a hole in an inner valence state, \textit{II)} the hole decays in a characteristic "transition time" and, due to electronic correlations, this can lead to excitation of a second electron to a bound excited state, hereafter called "shake-up" state, \textit{III)} the NIR pulse extracts the excited electron, hence doubly ionising the molecule. A simplified representation of this scenario is pictorially illustrated in Fig.~\ref{fig2}. As anticipated, the removal of an electron from a correlated state may result in a non-stationary charge distribution that rapidly evolves in time (charge migration). We could already speculate that the initiated charge migration process determines the optimal time window for an increased absorption of the NIR probe pulse. Moreover, the creation of the dication can only take place after the shake-up process has occurred but before the excited molecular cationic state relaxes via non-adiabatic couplings.


To corroborate our interpretation, we first evaluated characteristic shake-up times and searched for a peculiar one compatible with the experimentally observed time delay. The shake-up process illustrated in the level diagrams of Fig.~\ref{fig2}a is purely driven by electronic correlation (two-body Coulomb interaction), not accounted for in standard TDDFT simulations\cite{Cucinotta2012}. Nevertheless, a simple estimation of the shake-up transition time can be obtained with a rate equation approach: the initial statistical superposition of states created by the XUV pulse is calculated using ab-initio photoionisation probabilities, and Fermi's golden rule is used to obtain the shake-up rate due to the Coulomb interaction (see SI). Figure~\ref{fig2}(b) shows the characteristic shake-up transition times towards different bound excited orbitals (Kohn-Sham (KS) orbitals obtained with DFT ground-state calculations).
Most of the values are of the order of a few hundreds attoseconds except for three states, one of which (the LUMO+6 indicated in green) is 2.5~fs, very close to the experimentally observed delay of the parent dication formation. Interestingly, the energy of this orbital is in the window of two-photon ionisation from the NIR pulse, which is in agreement with the above-mentioned experimental observation that two NIR photons are required for the creation of the dication.

While the rate-equation approach is intuitive and it provides a clear physical explanation of the experimental findings, it is an overly simplified description since it treats electronic correlations in first order perturbation theory, in a non-dynamical fashion and lacks a first principles description of the XUV ionisation process. More refined and independent ab-initio calculations are required to further validate our interpretation and provide a predictive framework to address similar phenomena in other molecules. To this end, we performed many-body time-dependent simulations from first-principles to take into account both the electron dynamics triggered by the XUV photoionisation and the absorption of a delayed NIR pulse. By solving the equations of motion for the non-equilibrium Green's function and using an efficient propagation scheme based on the generalised Kadanoff-Baym ansatz, we can obtain an accurate and controlled treatment of shake-up processes\cite{Perfetto2018,Covito2018} and describe the light-molecule interaction from first-principles using laser pulses with the same characteristics as in the experiment (see SI). It is important to point out that our simulations are well suited to describe the electron dynamics but do not take into account the nuclear dynamics and therefore non-adiabatic effects (e.g conical intersections). From the calculations, we first extract the orbital-resolved occupations that are reported in Fig.~\ref{fig3} (a). When only the XUV photoionisation is considered, we can confirm the results obtained with the rate equations: while most of the states exhibit sub-femtosecond rise-times, the LUMO+6 occupation (shown in green) rises over several femtoseconds due to a slow shake-up process. Fig.~\ref{fig3} shows the integrated time-dependent electron density more than 3~{\AA} away from the molecular plane (panel (b)) and snapshots of the change in electron density (panel (c)). As it can be observed from this figure, a significant electronic \emph{charge inflation} builds up over a few femtoseconds. The spatial distribution of this density variation resembles the one of the LUMO+6 orbital (see SI), highlighting its dominant role in the overall electron dynamics. From Fig.~\ref{fig3}(b) it could be seen that the integrated electron density rapidly increases in the first 3~fs and therefore we could argue that this rapid delocalisation far from the molecular plane plays a dominant role in the delayed absorbtion of the NIR probe pulse for the efficient creation of the stable dication.
Our simulations also indicate that LUMO+6 can only be accessed when ionisation is triggered by an XUV pulse polarised perpendicularly to the molecular plane. Therefore, the relative orientation between the molecule and the attosecond pulse can potentially be exploited to achieve more refined control over the ionization process.

\begin{figure}[bh!]\centering
\includegraphics[width=0.7\textwidth]{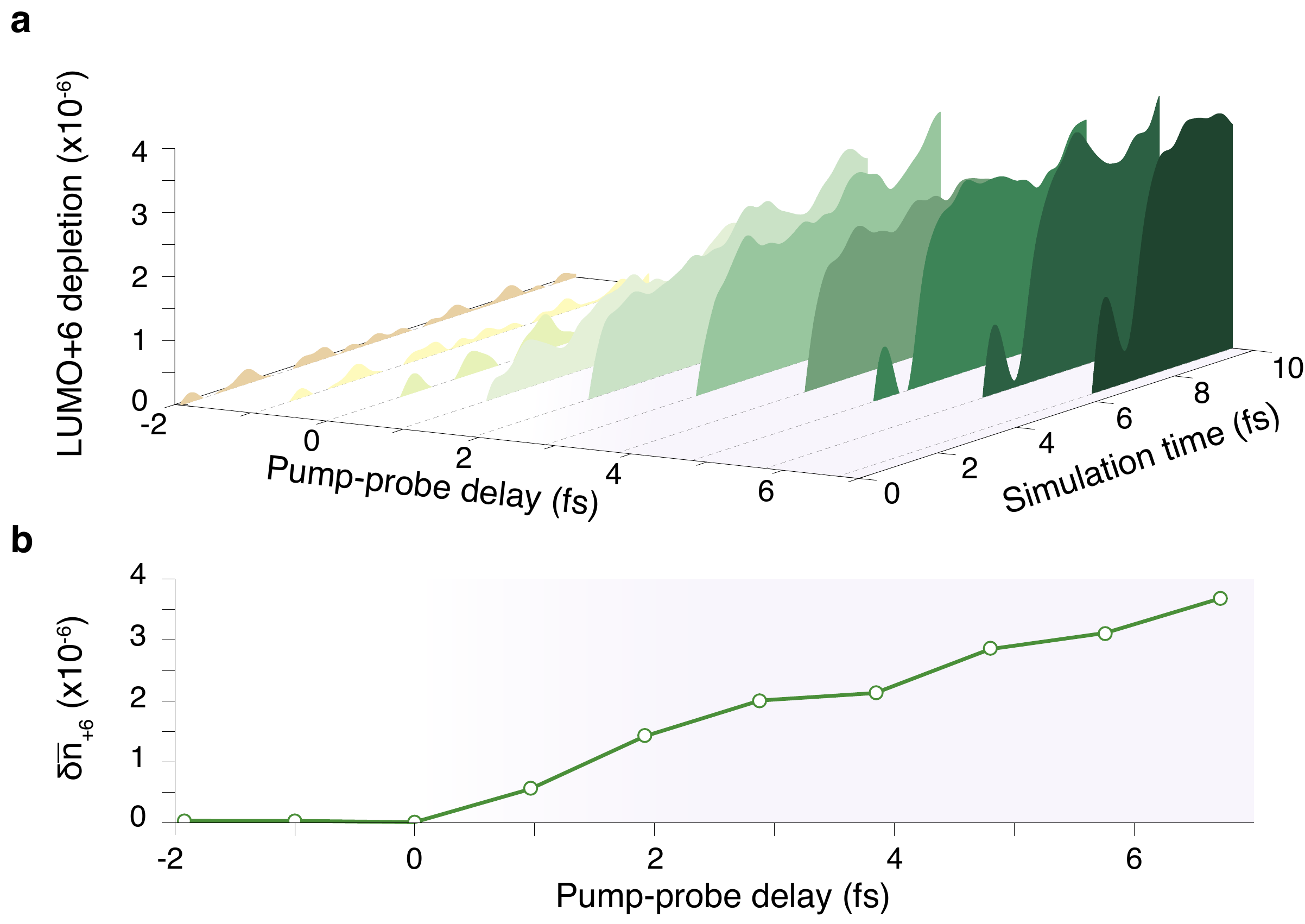}
\caption{\label{fig4}
\textbf{Role of the NIR probe pulse.}
(a) Temporal evolution of the LUMO+6 state depletion due to the combined action of XUV and NIR pulses as a function of the delay. The depletion shows a significant onset in the window of 2--4~fs and it keeps increasing with larger delays as shown in (b) where we report the state depletion averaged in a 1~fs window after the NIR pulse. The trend reported in (b) reproduces, remarkably, the one of the time-dependent yield measured for the adenine dication. The results in panels (a) and (b) are based on the non-equilibrium Green's function method.
}\end{figure}

Finally, we have included the NIR absorption in the simulation and calculated the time-resolved NIR-induced depletion, i.e. population reduction, of the LUMO+6 (Fig.~\ref{fig4}(a)), for different pump--probe delays. The depletion shows an onset in the window of 2--4~fs and increases with larger delays. To clarify the delay dependence of the depletion, we present the average depletion (over a 1 fs window following the NIR pulse) as a function of pump-probe delay in Fig.~\ref{fig4}(b). The trend reproduces, remarkably, the one measured for the adenine dication yield (green solid line in Fig. ~\ref{fig2}).
We point out that only the LUMO+6 state is characterised by this slow onset (other states in SI) and we can therefore conclude that this peculiar shake-up dynamics -- resulting in the out-of-plane charge migration mechanism -- explains why the NIR pulse has to be optimally delayed in order to produce the stable dication. We note that our model cannot reproduce the 24-fs exponential decay observed in the time-dependent dication yield, since non-adiabatic couplings have been neglected. We presume however that this decay is a signature of the dephasing induced by the recently depicted electron-phonon like coupling occurring in correlation bands for large molecules\cite{Herve2020}.

To summarise, our theoretical calculations singled out a special shake-up state (LUMO+6) leading to an out-of-plane charge migration mechanism, which mediates the observed delayed creation of stable doubly charged adenine. The peculiarity of this state can be attributed to the following characteristics: I) it has a few-femtoseconds shake-up time, compatible with the experimentally observed delay in the dication formation, II) it is a delocalised excited state that extends away from the molecular plane and III) it couples very efficiently to the NIR pulse. Our findings not only indicate that the delayed creation of the dication is a valuable probe for a many-body effect, but also that by precisely timing a NIR control pulse one could take advantage of a correlation-driven charge redistribution to prevent dissociative relaxation.

\spacing{1.6} 

\spacing{1.7} 
Real time mapping of the many-body effects in ionised biochemically relevant molecules is the first (crucial) step towards a control of the molecular reactivity at the electronic time scale. Here we have not only characterised the intrinsic time for a shake-up process to occur in ionised adenine but also the resulting \emph{out-of-plane charge migration}, neither of which to our knowledge has been measured in real time for any polyatomic molecule before. A key aspect here is that the many-body effect mediate the efficient absorption of a properly delayed NIR control pulse, leading to the creation of intact and stable doubly charged adenine. 
Our findings demonstrate that extreme time resolution is required to act promptly after molecular ionisation and take advantage of the non-equilibrium charge distribution (before non-adiabatic effects take place) to achieve control over the molecular dissociation. By complementing the experiments with covariant detection of electrons and ions and with the support of more advanced time-dependent many-body methods, including the nuclear motion, we could potentially obtain a direct mapping of this electronic redistribution and target more specifically new stabilisation pathways for a wide range of polyatomic molecules.

\clearpage 
\section*{References}


\begin{addendum}
\item[Data availability] The data that support the findings of this study are available from the corresponding author upon reasonable request.
\item[Code availability] The program code used to treat experimental data is available from the corresponding author upon reasonable request. The simulations of electronic dynamics were done with previously described packages\cite{Perfetto2018}.
\item[Funding] F.Ca. acknowledges support from the European Research Council under the ERC-2014-StG STARLIGHT (Grant Agreement No. 637756). F. Ca and A. R. acknowledge support from the Deutsche Forschungsgemeinschaft (DFG, German Research Foundation) – SFB-925 – project 170620586 and the Cluster
of Excellence Advanced Imaging of Matter (AIM). F.L. and V.W. acknowledge the Fonds de recherche du Qu{\'e}bec -- Nature et technologies (FRQNT) and the National Science and Engineering Research Council (NSERC). V.W. acknowledges support from the Vanier Canada Graduate Scholarship (Vanier CGS) program. S. L. acknowledges support from the Alexander von Humboldt foundation. A. R. acknowledge financial support from the European Research Council(ERC-2015-AdG-694097). The Flatiron Institute is a division of the Simons Foundation. G.S. and E.P. acknowledge EC funding through the RISE Co-ExAN (Grant No. GA644076), the European Union project MaX Materials design at the eXascale H2020- EINFRA-2015-1, Grant Agreement No. 676598, Nanoscience Foundries and Fine Analysis-Europe H2020-INFRAIA-2014-2015, Grant Agreement No. 654360 and Tor Vergata University for financial support through the Mission Sustainability Project 2DUTOPI. J. B. G. acknowledge support from the EPSRC (UK) grant number EP/M001644/1.
\item[Author contributions] 
S.L., E.P, G.S., L.P., F.L., M.N., A.R. and F.Ca. supervised the project.
E.M., V.W., M.G. and M.C. performed the measurements, with F.F. and J.G. contributing resources. 
S.L., F.Co., E.P. and G.S. performed the simulations.
E.M., S.L., F.Co., H.H., U.D., A.T. and F.Ca. wrote the manuscript.
\item[Acknowledgements] The authors would like to acknowledge A. Stolow for the fruitful discussion.
\item[Competing interests] The authors declare no competing interests.
\item[Supplementary information] is available online. (Supplementary Text, Figures S1 to S18, Tables S1 to S2 and References 30--49.)
\item[Correspondence and requests for materials] should be addressed to F.Ca.
\end{addendum}

\end{document}


\vspace*{1em} 
\title{Supplementary Information}  
\maketitle
\thispagestyle{plain} 

\onecolumngrid
\vspace{-1.3em} 
\noindent{This PDF file contains: Supplementary Text, Figures S1 to S18, Tables S1 to S2 and References 34 to 52.}
\vspace{0.6em}
\twocolumngrid

\nocite{Cederbaum1999,Calegari_2015,Drescher2002,Hanna2017,Averbukh2011}
\nocite{Kuleff2013,Kuleff2014,Remacle2006,Nisoli_2017,Calegari2014}
\nocite{Lara-Astiaso2018,Stolow2013,Jenkins2016,Polyak2018}
\nocite{Lara-Astiaso2016,Marciniak2019,Herve2020,Corkum1994}
\nocite{Sola2006,Pilling2007,Alvarado2007,Bredy2009,Burgt2015}
\nocite{Galbraith2017,Marques2003,Castro2006,Andrade2015}
\nocite{runge_density-functional_1984,bertsch_real-space_2000} 
\nocite{Alonso:2008fz,Cucinotta2012,Perfetto2018,Covito2018}

\section{Experiment}

\subsection{Experimental set-up and data acquisition}
\label{ch:acquisition}

\begin{figure}
\hspace{-4mm}\includegraphics[width=1.05\columnwidth]{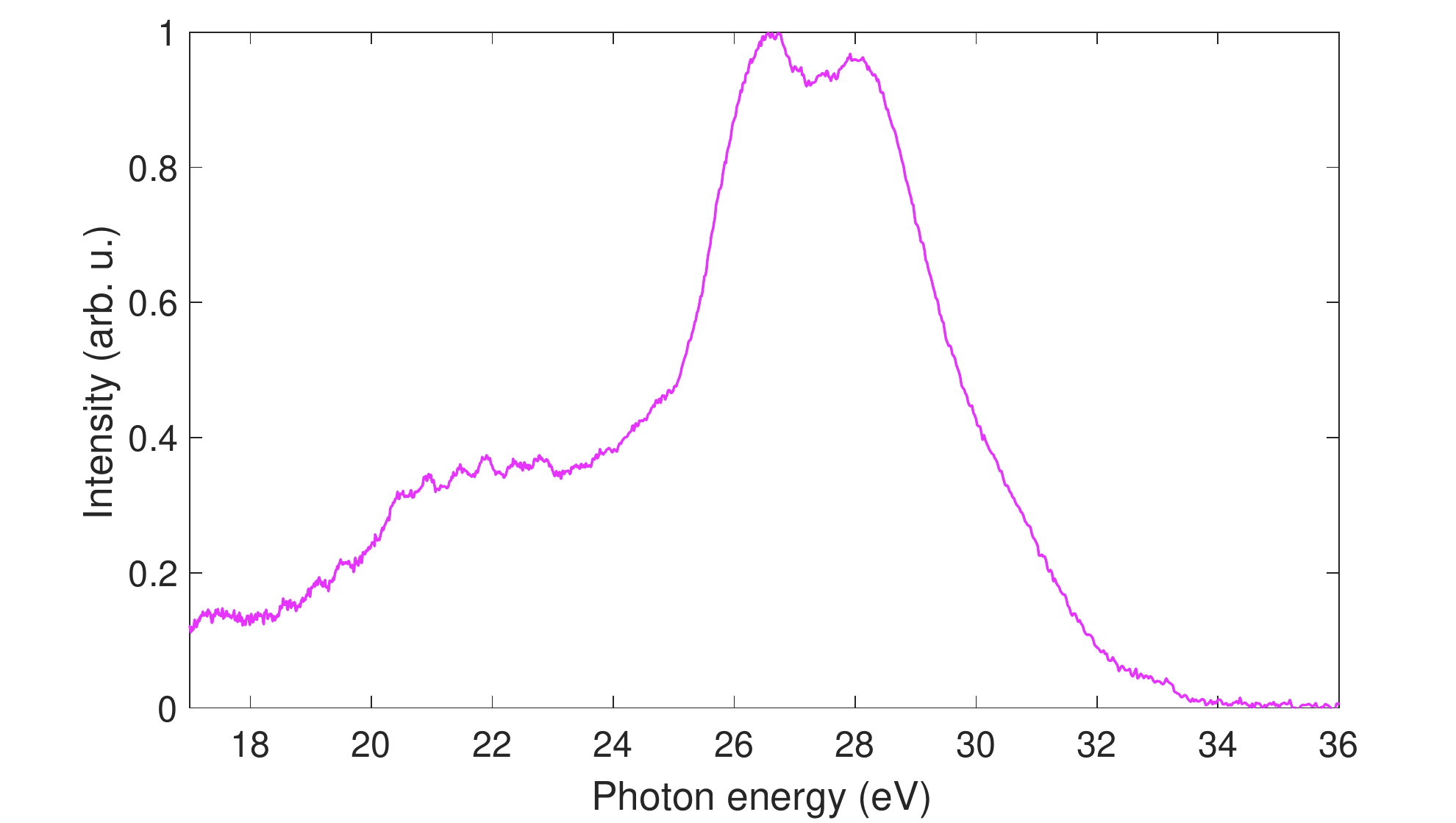}\vspace{-1.5mm}
\caption{\label{figS-XUV}
\textbf{XUV spectrum of isolated attosecond pulse generated in krypton.}
}\end{figure}

Carrier-envelope-phase stable laser pulses with 4\;fs duration, 2.5\;mJ energy and a central wavelength of approximately 750\;nm are used to drive the attosecond pump--probe setup \cite{Calegari2014,Mansson2017,Seiffert2017}. The beamline is based on a Mach--Zehnder-type interferometer where one arm (70\;\% of the initial energy) is used for XUV generation while the other arm provides the NIR probe pulse at  adjustable delays. Collinear recombination of the two beams is achieved with a drilled mirror reflecting the IR and transmitting the XUV in the central hole. Sub-300 as XUV pulses are generated in krypton or xenon (depending on the desired cut-off energy) by polarisation gating\cite{Sola2006}. A typical spectrum generated in krypton is shown in \autoref{figS-XUV}. A 100-nm-thick aluminium filter is used to remove the residual NIR (as well as harmonics below 15\;eV) from the XUV arm. The attosecond interferometer is actively delay-stabilised with a residual RMS of 20\;as.

Adenine powder (Sigma-Aldrich, $>$99\;\%) is evaporated at 463~K in a resistively heated stainless steel oven with a flow of helium acting as carrier and buffer cooling gas. 
According to vibrational spectroscopy on adenine from a similar source, the lowest-energy tautomer (9H amino) dominates \cite{Pluetzer2001}. 
The jet passes through a 1-mm diameter skimmer and down to the optical interaction region of a standard velocity map imaging spectrometer \cite{Eppink1997} operated as a time of flight mass spectrometer. The voltages of the microchannel plate (MCP) and phosphor screen are gated to avoid detecting the helium gas at short times of flight. 

Pump--probe data were acquired by scanning the delay in alternating directions over multiple traces (7 in the main dataset used, 5 for the atomic time reference scan). Each trace was normalised to the total yield to account for a slow sample depletion. These multiple traces have been used to calculate the errorbars shown for the experimental signal, as a shaded grey area in Fig.~1b and \autoref{figS-timeref}. 

\subsection{Estimation of fragmentation probability}
\begin{figure*}[t!] 
\includegraphics[width=1.35\columnwidth]{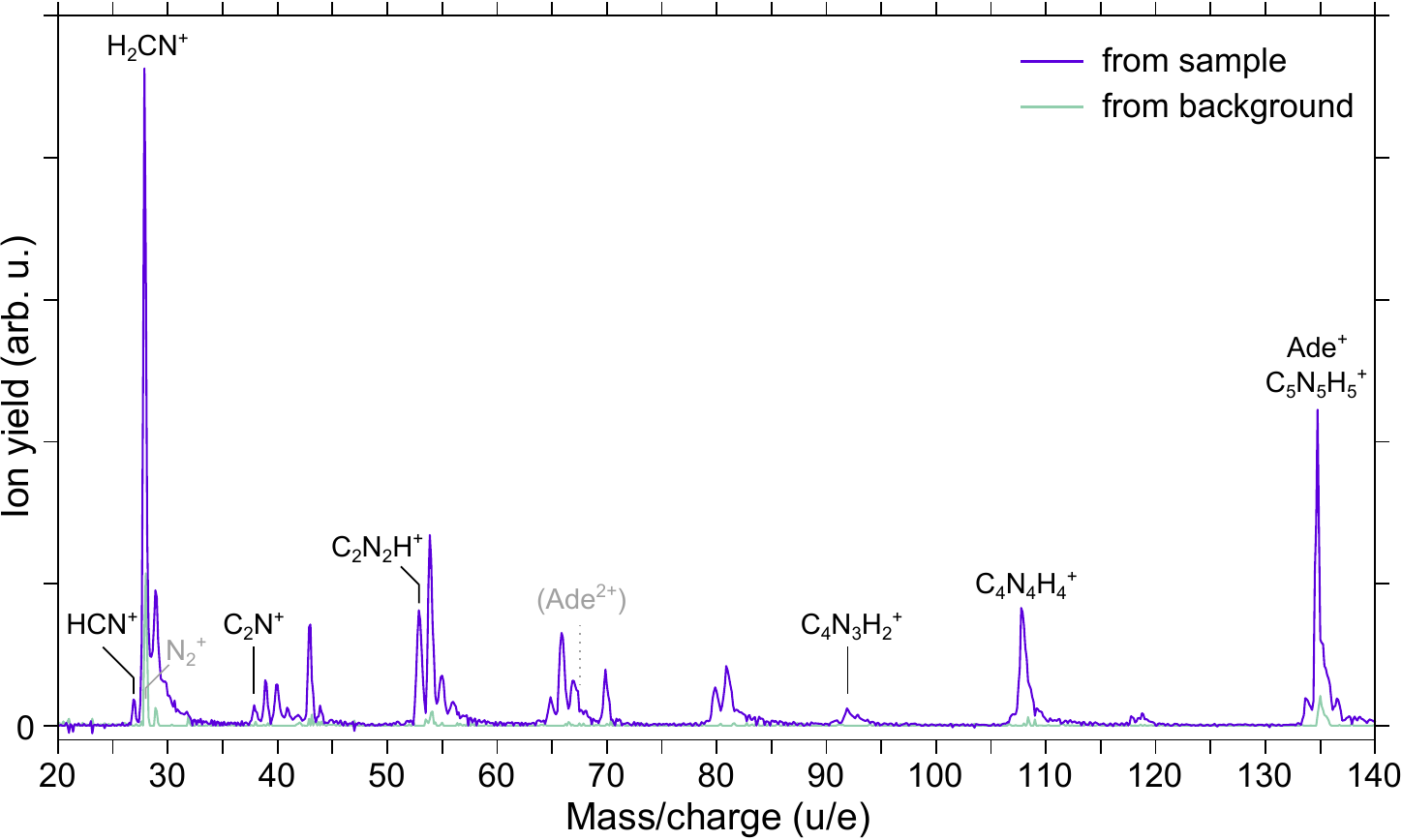}\vspace{-1.5mm}
\caption{\label{figS-dication-mass-separation}
\textbf{Separation of background and adenine contributions to the XUV mass spectrum.}
Of the raw peak at 28~u/e, 16\% was deduced to be N$_2^{+}$ from the background instead of HCNH$^{+}$ from adenine.
The dotted line indicates where Ade$^{2+}$ would appear (67.5 u/e), but it is essentially absent without NIR probe (compare \autoref{figS-dication-appearance}).
}\end{figure*}

To estimate the fraction of XUV-ionized adenine molecules that dissociate we can compute the fraction of ions with lower mass than adenine, within the mass range detectable with the gated MCP operation. Background gases could however exaggerate this fraction, for instance by nitrogen gas having the same molecular mass as the H$_2$CN$^+$ fragment (at 28~u/e in \autoref{figS-dication-mass-separation}). We therefore separate the XUV-only mass spectrum into sample and background components using the following generic method:

Two mass spectra, $y_\mathrm{full}(m)$ and $y_\mathrm{reduced}(m)$, need to be recorded with identical conditions (e.g. XUV spectrum and residual gas density) but with different density of the adenine sample, achieved by changing the oven temperature. 
We use a scaling factor $0 < x < 1$ to represent the remaining adenine density in $y_\mathrm{reduced}$ with respect to the full density in $y_\mathrm{full}$:
\begin{equation}
\begin{cases}
y_\mathrm{full}(m)   &= y_\mathrm{background}(m) + y_\mathrm{sample}(m)
\\
y_\mathrm{reduced}(m)&= y_\mathrm{background}(m) + x\, y_\mathrm{sample}(m) \label{eq:separation-system}
\end{cases}
\end{equation}
To determine the scaling factor, a region of the mass spectrum is defined as background-free, e.g. the region covering the parent cation.
Solving equation \eqref{eq:separation-system} for $x$ in the background-free region, where $y_\mathrm{background}(m) = 0, m \in M_\mathrm{parent}$, gives
\begin{gather}
x = \left.
		\sum_{m \in M_\mathrm{parent}} y_\mathrm{reduced}(m)
	\right/
		\sum_{m \in M_\mathrm{parent}} y_\mathrm{full}(m)
\\
y_\mathrm{background}(m) = \frac{y_\mathrm{reduced}(m) - x y_\mathrm{full}(m)}{1 - x}
\label{eq:separation-background}\\
y_\mathrm{sample}(m) 
= y_\mathrm{full}(m) - y_\mathrm{background}(m)
\label{eq:separation-foreground}
%
\end{gather}

\begin{figure*}[t!] 
\includegraphics[width=1.35\columnwidth]{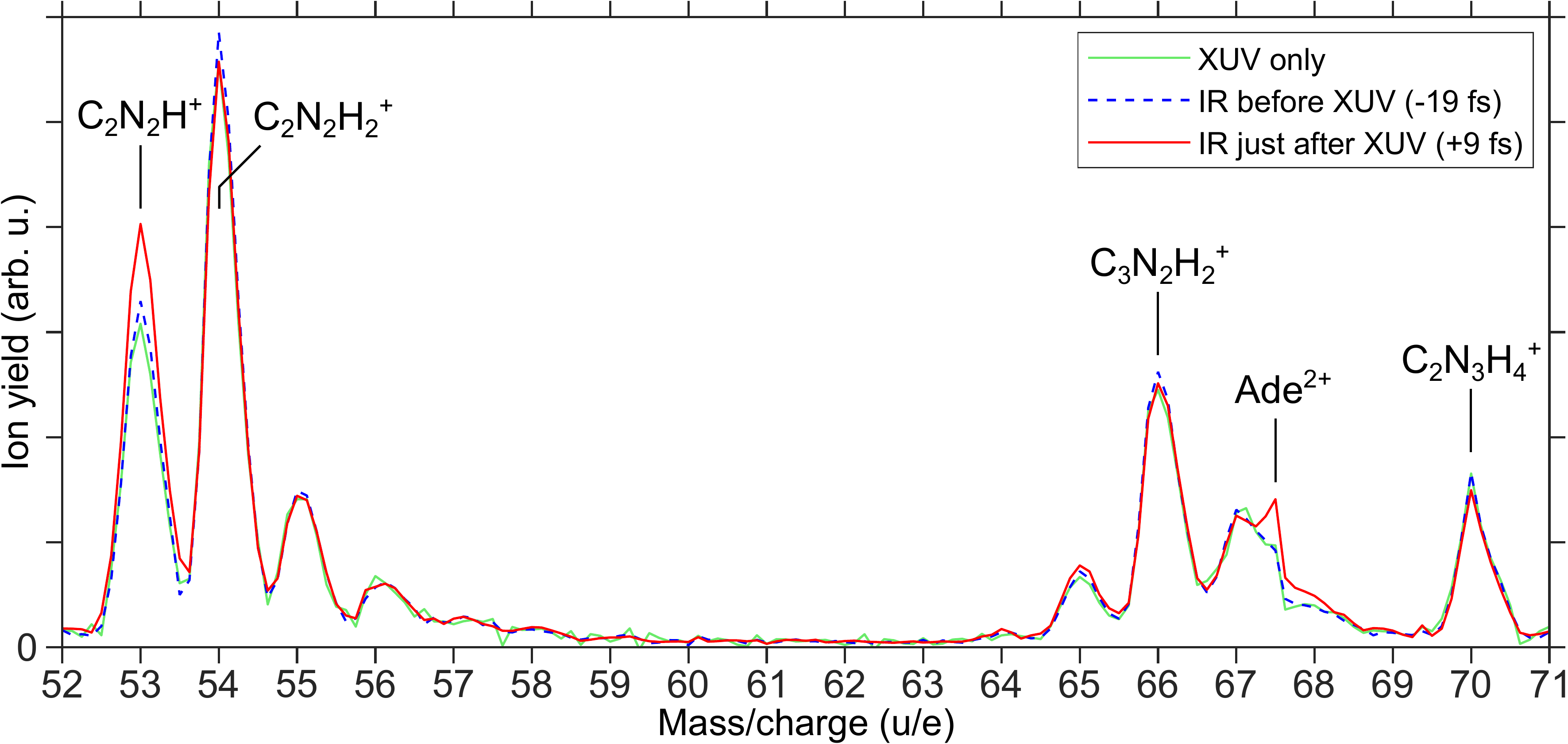}\vspace{-1.5mm}
\caption{\label{figS-dication-appearance}
\textbf{Appearance of the adenine dication.}
Zoom-in on the mass spectrum where the dication clearly appears at a small positive NIR delay but is negligible by the XUV alone or with the NIR at a negative delay.
}\end{figure*}

As the separation is based on the difference between two mass spectra, the output contains some noise and artefacts. A simple correction that was made was to require both the background and sample signals to be non-negative. After evaluating \eqref{eq:separation-background}, $y_\mathrm{background}$ is updated at individual mass/charge-bins $m$ to ensure that $0 \leq y_\mathrm{background}(m) \leq y_\mathrm{full}(m)$. The latter inequality ensures that also the final $y_\mathrm{sample}$ remains non-negative when computed as in \eqref{eq:separation-foreground}.

The resulting separated adenine and background mass spectra are shown in \autoref{figS-dication-mass-separation} for masses above 20~u/e.
In this range, the adenine mass spectrum consists of 81\% fragments and 19\% parent cation.
The background gas removal procedure was not applied to pump--probe data, as it was introduced for the sole purpose of estimating the total fragment yield.



\subsection{Adenine dication formation as a function of the XUV photon energy}

We show that the stable adenine dication yield is maximised when the NIR probe pulse is sent at a small but nonzero delay after the XUV pump. As a complement to the delay-dependence curves of individual ions in Fig.~1b we show mass spectra at negative and positive NIR delay in \autoref{figS-dication-appearance}, together with the XUV-only case. Since sending the NIR pulse before the XUV pulse gives the same result as the XUV only case, we conclude that there is no NIR-pump--XUV-probe contribution.

At 67.5~u/e, where the dication appears, there is also a background contribution due to the width of the adjacent fragment at 67~u/e. Since the unambiguous part of the 67~u/e peak shows no delay-dependence, this additional constant background will not have any influence on the results concerning delay-dependence (such as Fig.~1b). Estimating the absolute yield of the dication, however, requires the background to be subtracted. From the perturbed curve shape, we estimated a 0.2~\% production of adenine dication with XUV only, which we consider almost negligible. 

\begin{figure}[b!]
\includegraphics[width=0.82\columnwidth]{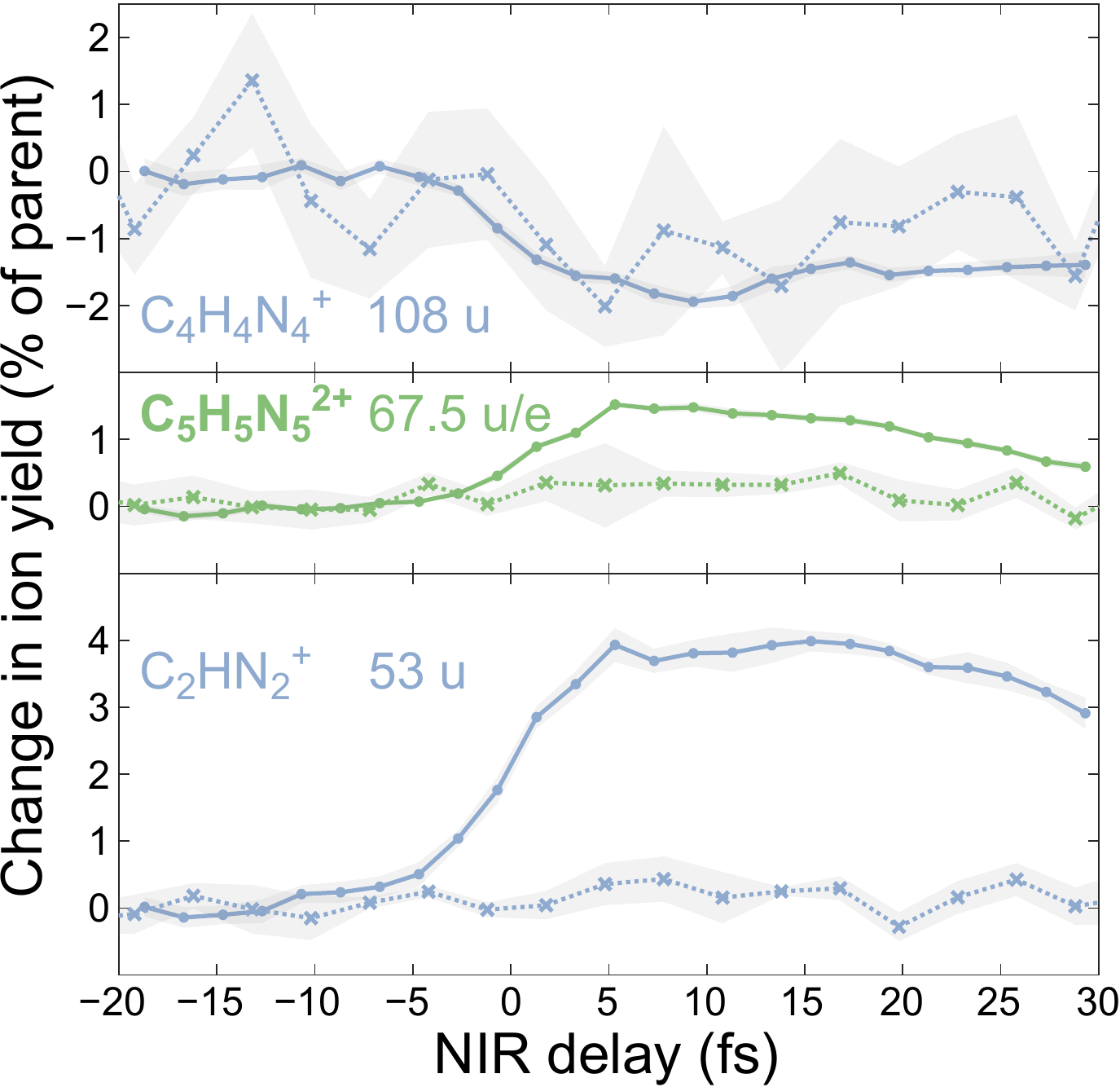}
\caption{\label{figS-XUV-cutoff}
\textbf{Pump-probe results for different XUV photon energies.}
Pump--probe scan signals are shown in solid lines for the main scan (35 eV cut-off) and dotted lines for indium-filtered HHG from xenon (17 eV cut-off).
}\end{figure}

In \autoref{figS-XUV-cutoff} we examine the effect of limiting the XUV photon energy on the pump--probe dynamics. The three panels show the NIR-induced change in the yields of the two cationic fragments with the largest absolute change as well as the adenine dication. Solid lines correspond to the normal case where the XUV photon energy of the attosecond pulses extends from 15 to 35\;eV, employing XUV generation in Kr and an aluminium filter.
Dotted lines show a scan where the XUV generated in Xe is limited to photon energies below 17\;eV by an indium filter (about 5.2\;eV below the dication ground state). In this last case, the absorption of one XUV photon and a few NIR photons is not able to produce any stable dication. Consequently, we could conclude that population of the special state (identified as responsible of the stabilisation process) can be only observed if ionisation occurs using the higher-energy part of the XUV spectrum. We also note that with the reduced cut-off, many fragments disappear, even the largest NIR-induced absolute step for the 53~u/e fragment. For 108~u/e there may still be a negative step, but close to the noise level.


\subsection{Number of NIR photons in the studied process}
\label{section:number-of-NIR-photons}

\begin{figure}[b]
\vspace{-1mm}\includegraphics[width=1\columnwidth]{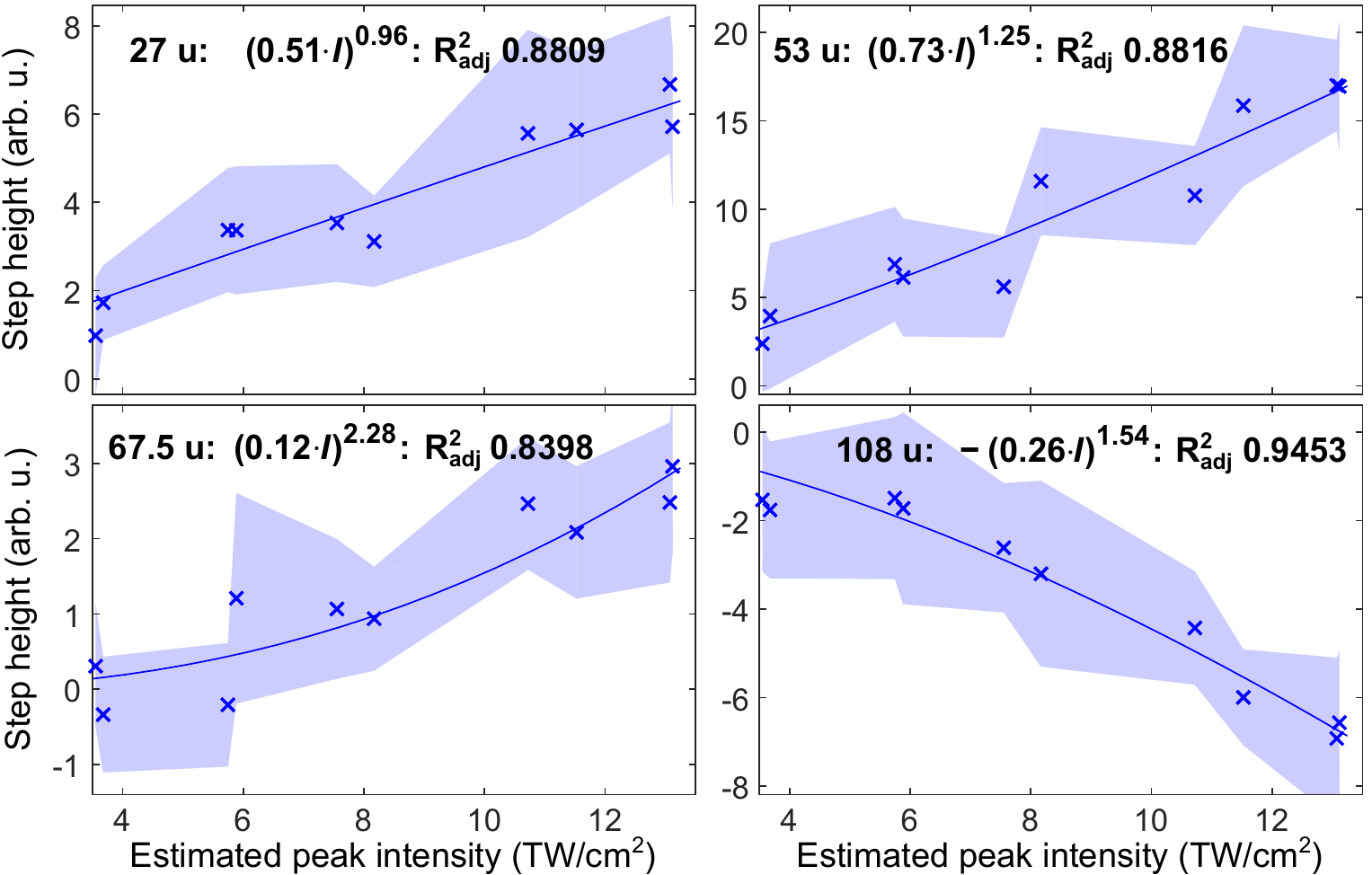}
\\\vspace{3mm}
\includegraphics[width=0.84\columnwidth]{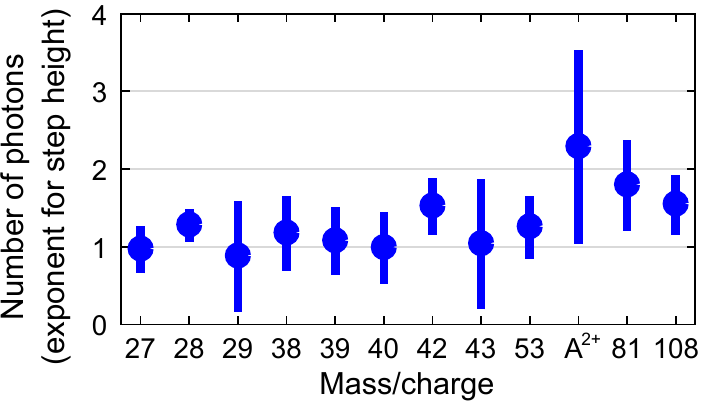}
\caption{\label{figS-NIR-photons}
\textbf{Scaling of step heights with NIR intensity.}
The delay-dependent step height scales linearly with NIR intensity for several small fragments but approximatively quadratically for the adenine dication.
}\end{figure}

A sequence of short scans were made over a range of NIR intensities from approximately 7$\times 10^{12}$ to 1.4$\times 10^{13}$~W/cm$^2$ to estimate the number of NIR photons involved in the XUV+NIR process(es) producing each detected ion. We first extracted the step height, $h_m$, of the pump-probe signal for each ion (at mass/charge) at different NIR intensities. We then fitted the NIR-intensity-dependent data with the power law $h_m(I) = (c_m I)^{n_m}$, where $I$ represents the NIR intensity and $n_m$ represents the number of NIR photons driving the process in addition to the single XUV photon (see top panels of \autoref{figS-NIR-photons}).
Although the interpretation of $n_m$ is straightforward for the parent cation and dication, one must for a fragment ion see it as a net result of the NIR probe's competing enhancement (via dissociation of the parent and larger fragments) and depletion (via further dissociation) of the signal.
We acknowledge that there is some uncertainty (possibly 25~\%) in the absolute NIR intensities in \autoref{figS-NIR-photons}, but a linear rescaling of the intensities would not affect the determination of the nonlinearity order $n_m$, only the coefficient $c_m$.

The results of the fitting procedure are reported in the bottom panel of  \autoref{figS-NIR-photons}: the step heights of the small fragments scale approximately linearly while the adenine dication's step height scales approximately quadratically with the NIR intensity. This result suggests that 2 NIR photons are required in the probing step (after shake-up) of the process we discuss in the main manuscript, yielding a non-dissociated doubly charged adenine dication.


\clearpage

\subsection{Fitting of the pump--probe signals}

When an initial state populates the probed state at rate $1/\tau_1$, and the probed state decays with the rate $1/\tau_2$, the general solution to the rate equations allows the population of the probed state to be expressed as:
\begin{equation}
N(t) =
\begin{cases}
    N_1 \frac{1}{1 - \tau_1/\tau_2} \left(  \mathrm{e}^{-t/\tau_2} - \mathrm{e}^{-t/\tau_1} \right) &,\; t > 0 \\
    0 &,\; t < 0
\end{cases}
\label{eq:curve-population}
\end{equation}
under the condition that $0 \leq \tau_1 < \tau_2$, i.e. the state population will rise to a positive peak and then decay. 
Here, $t$ is the pump--probe delay axis of the scan, positive when the NIR pulse comes after the XUV attosecond pulse. An arbitrary coefficient $N_1$ is included for generality.

To model experimental pump--probe scans, the population function should be convoluted with an instrument response function and the temporal overlap of the XUV and NIR pulses needs to be defined with a $t_0$-parameter.
We allow for a constant background level $b$ and name the step height parameter $h$.
Since convolution is a linear operation, we can convolve the two terms in \eqref{eq:curve-population} separately and express the complete curve model as
\begin{equation}
f(t) =  b + \frac{h}{1-\tau_1/\tau_2}
\big( 
    g_{\tau_2}(t - t_0)  
  - g_{\tau_1}(t - t_0)  
\big).
\label{eq:curve-model}
\end{equation}
Here, $g_{\tau}(t)$ represents the convolution of a single exponentially decaying step function with a Gaussian instrument response function\cite{Galbraith2017} of full-width at half-maximum $W = 2\sqrt{2 \ln 2} \sigma_\mathrm{IRF}$. The convolution is evaluated using the expression\cite{Lacoursiere1994}
\begin{align}
g_{\tau}(t) = &\frac{1}{2}
    \mathrm{erfc}\left(
    \frac{\sigma_\mathrm{IRF}}{\tau \sqrt{2}}
    - \frac{t}{\sigma_\mathrm{IRF} \sqrt{2}} 
    \right)\nonumber\\
    &\times \mathrm{exp}\left(-t/\tau + \sigma_\mathrm{IRF}^2/(2\tau^2) \right)
\end{align}
and normalized $(\int_{-\infty}^{\infty} g_{\tau}(t) \mathrm{d}t = \tau$) such that the step height before convolution is 1. 

\begin{figure}[b]
\includegraphics[scale=0.49]{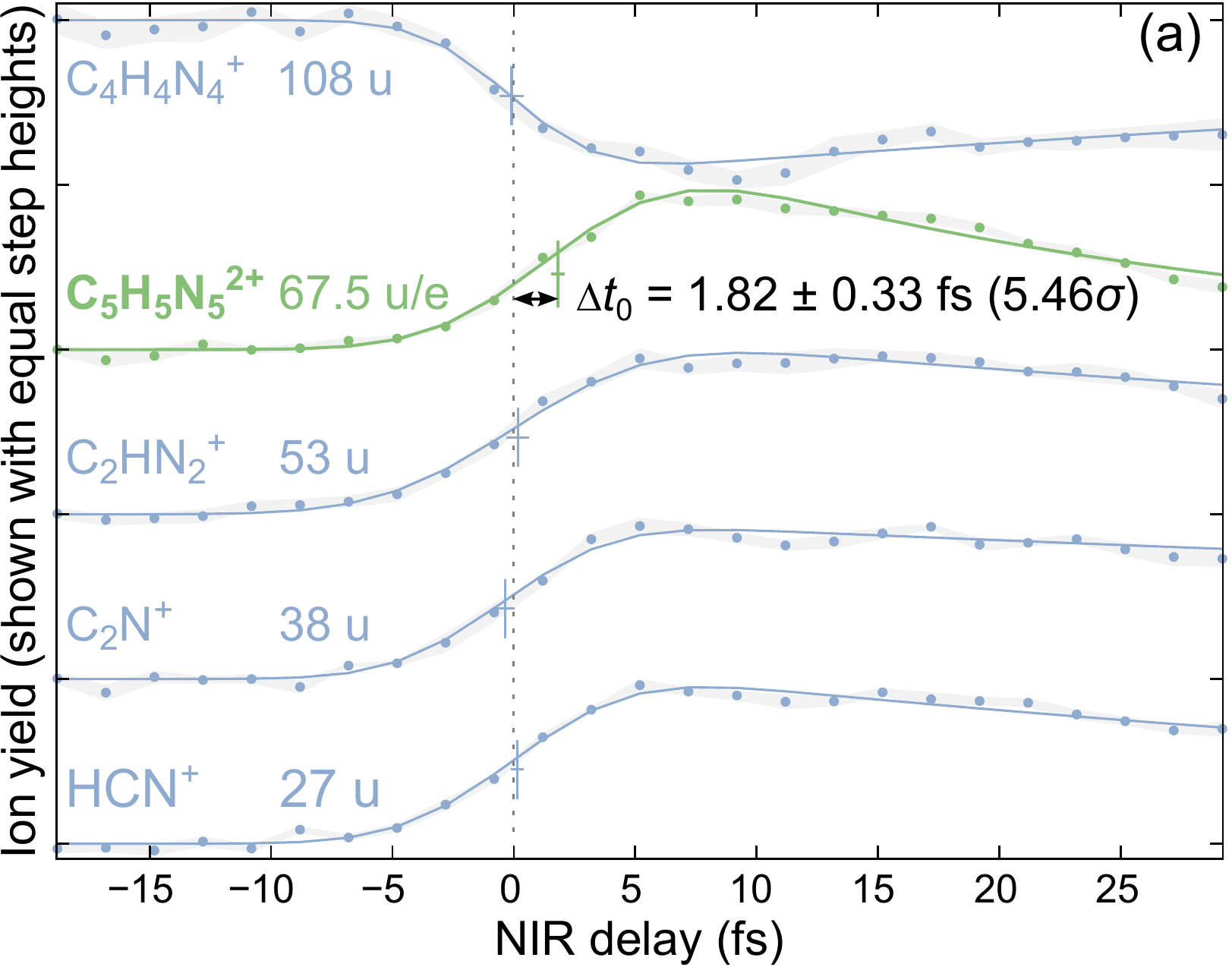}\\
\includegraphics[scale=0.49]{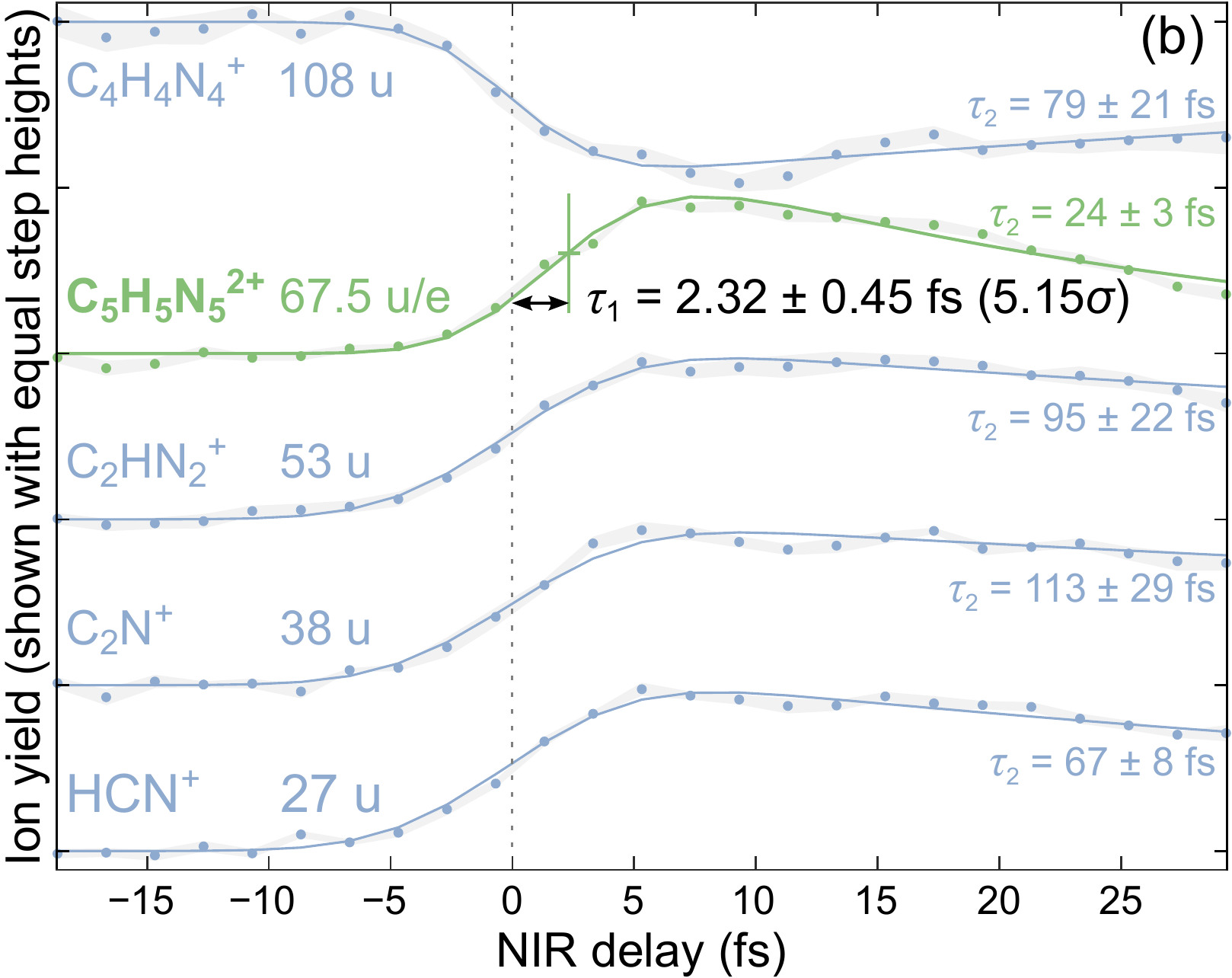} 
\caption{\label{figS-scan-variants}
\textbf{Pump--probe scan fitted using different approaches}.
(a) Simple step functions (having $\tau_1=0$) at individually variable $t_0$ were fitted without constraints. 
We find that the adenine dication (C$_5$N$_5$H$_5^{2+}$, 67.5~u/e) has a distinct delay with respect to the weighted $t_0$-average of the cationic fragments, and show the NIR-delay axis relative to the cation average.
(b) To obtain the more theoretically meaningful risetime of the adenine dication signal, $\tau_1$ was included in the function for the dication. 
The $t_0$-parameter is now common for all ions and defines the zero of the NIR-delay axis, and for robustness we use a common $W$-parameter fitted to $9.9\pm0.5$\,fs. The dication and the 108~u/e fragment where assigned as 2-photon processes for the best fit (in agreement with \autoref{section:number-of-NIR-photons}), meaning that their Gaussian width was $W/\sqrt{2}$ instead of $W$.
Panel (b) corresponds to Fig.~1b in the main article.
}\end{figure}

For an initial overview, we employed the simplest possible scheme where separate fits were made for each ion, with independent values for $t_0$ and $W$, and with simple step functions ($\tau_1 = 0$). From the result in \autoref{figS-scan-variants}(a), it is clear that a later $t_0$-parameter is needed for the dication.
We compute $\Delta t_0 = t_0(\mathrm{Ad}^{2+}) - \bar{t_0}(\mathrm{reference~cations}) = 1.82 \pm 0.33$\;fs. 
A classical time-difference like $\Delta t_0$ is however not physically meaningful. To obtain a result which can be compared with the theoretically extracted shake-up time, we introduced a risetime of the adenine dication signal, by using a variable $\tau_1$ in the function for Ade$^{2+}$. 
The $t_0$-parameter is now common for all ions and defines the zero of the NIR-delay axis, and for robustness we use a common $W$-parameter and perform a global fit of all the shown ions. The dication and the 108~u/e fragment where assigned as 2-photon processes for the best fit (in agreement with \autoref{section:number-of-NIR-photons}), meaning that their Gaussian width was $W/\sqrt{2}$ instead of $W$ since a \emph{two}-photon signal scales with the \emph{square} of NIR intensity envelope.
The results in \autoref{figS-scan-variants}(b) show a significantly nonzero risetime for the dication $\tau_1 = 2.32 \pm 0.45$\,fs, which we can now compare to the theoretically extracted shake-up time (see main text).



\subsection{Verification of zero time delay}
\label{text:atomic-time-reference}

\begin{figure}[b!]
\includegraphics[scale=0.49]{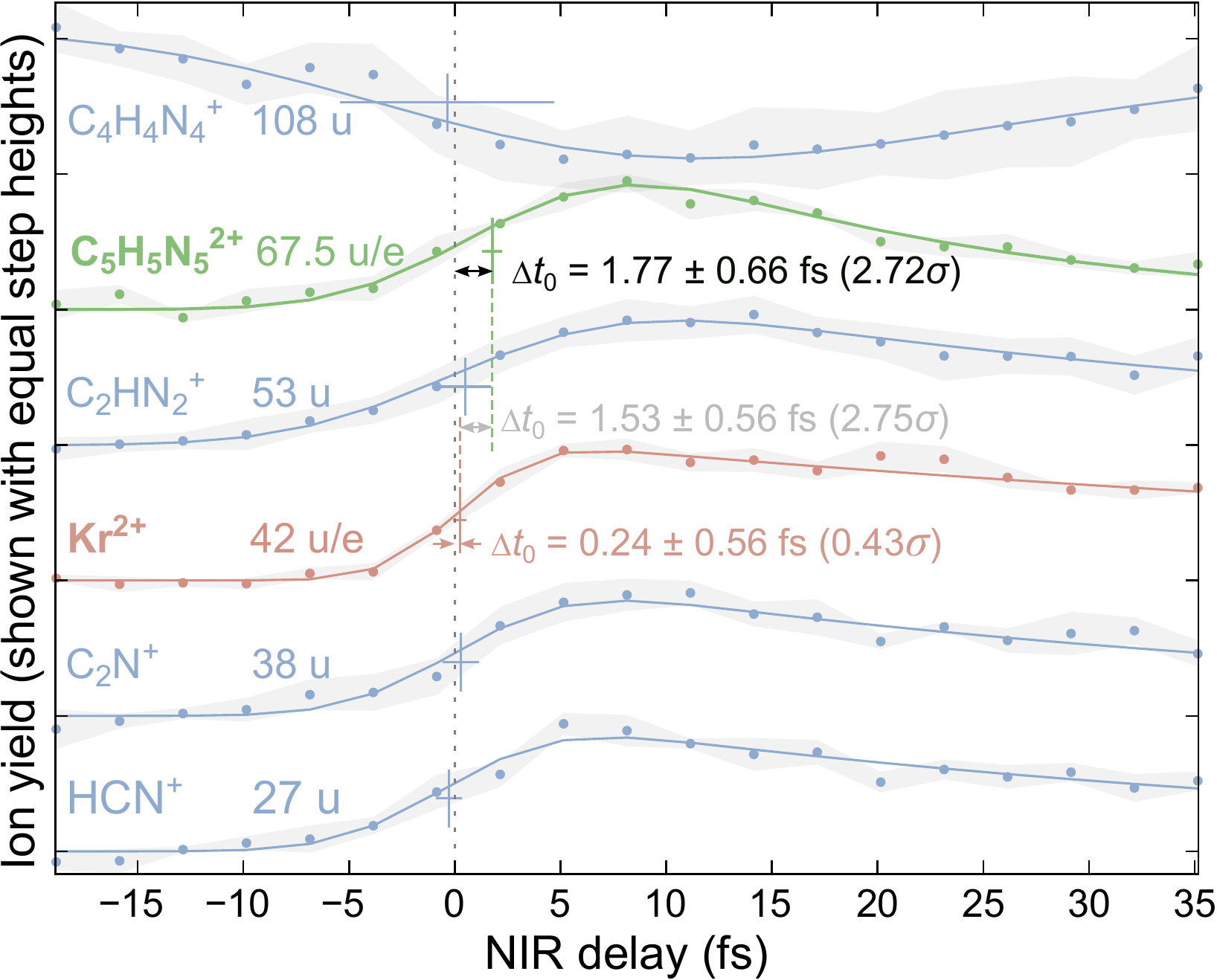}
\caption{\label{figS-timeref}
\textbf{Pump--probe scan with Kr$^{2+}$ (42~u/e) as atomic time reference}.
Three $\Delta t_0$ parameters, extracted using the first fitting procedure described in the previous section, are displayed: the difference between Ade$^{2+}$ and the reference cations (black text), the difference between Ade$^{2+}$ and Kr$^{2+}$ (grey text) and the difference between Kr$^{2+}$ and the reference cations (red text). Ade$^{2+}$ is significantly delayed while the cationic fragments are simultaneous with the Kr$^{2+}$ reference.
}\end{figure}

In our main analysis we discuss the delay of the adenine dication under the assumption that the steps of many cations occur directly at the temporal overlap of XUV and NIR pulses. Although this is is the natural assumption when so many fragments appear synchronised, one could still speculate that some common molecular motion is required before the NIR probe influences any cation fragment yield, shifting all steps to a positive NIR delay.
To rule this possibility out, we performed another experiment where a small amount of the atomic gas krypton was simultaneously injected into the vacuum chamber with adenine. The Kr$^{2+}$ yield in \autoref{figS-timeref} exhibits a sharp step (2-photon process meaning width parameter $W/\sqrt{2}$) for which any molecular dynamics can be excluded. 
For simplicity, we used only the first fitting procedure described in the previous section to extract a $t_0$-parameter for Kr$^{2+}$ to be compared with the other cationic fragments. The results of the fitting are shown in \autoref{figS-timeref}. We find that the Kr$^{2+}$ step almost coincides with the steps for the adenine fragments and that Ade$^{2+}$ is the only significantly delayed ion.

\FloatBarrier

\section{Theory}

\subsection{Equilibrium properties}

\begin{figure}[b!]\vspace{-3mm}
  \includegraphics[width=0.5\columnwidth]{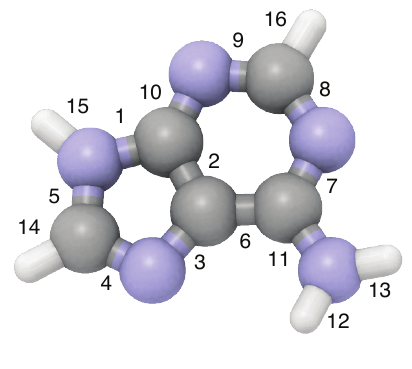}\vspace{-5mm}
  \caption{\label{fig:AdCartoon}\textbf{Adenine molecule bond numbering}} 
\end{figure}
\begin{figure}[b!]
\vspace{-1mm}\centering
\label{fig:bonds_go_charged}
\includegraphics[width=1\columnwidth]{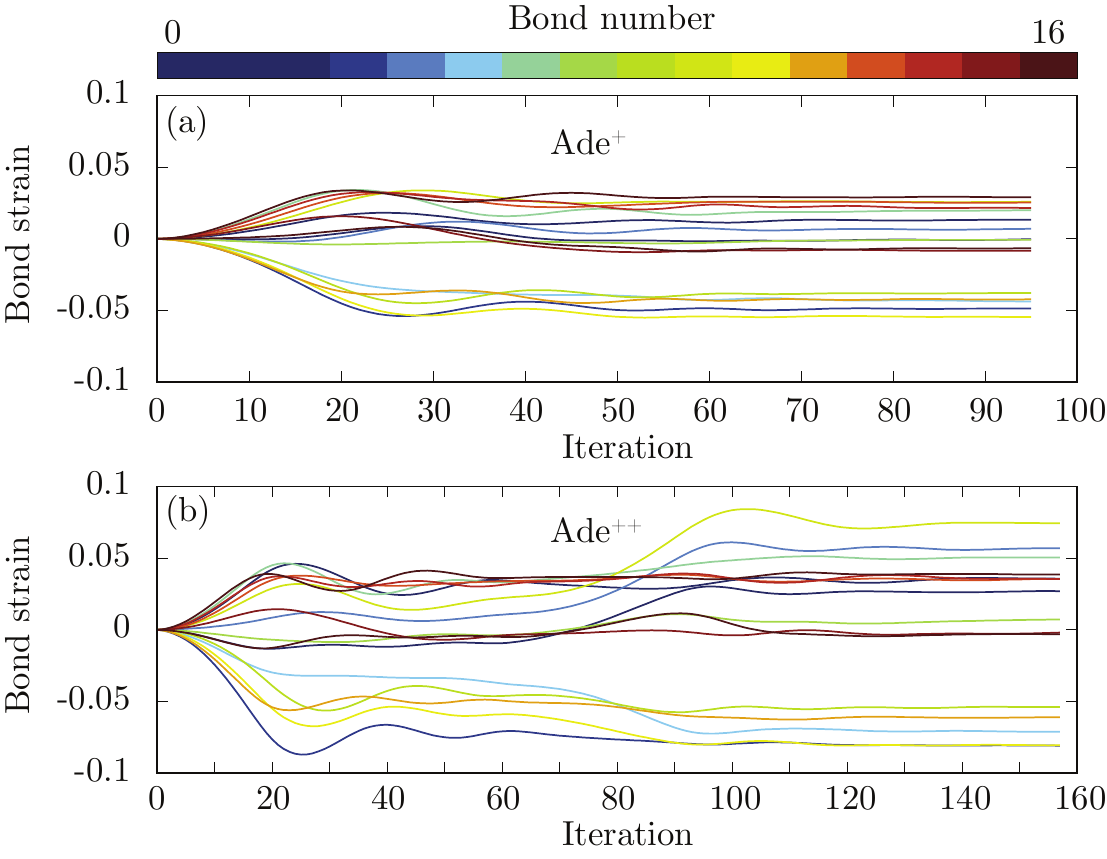}\vspace{-2mm}
\caption{{\bf Geometry relaxation}. relative variation of the bond length as a function of the optimisation steps for singly (a) and doubly (b) ionised adenine. The reference bond length is the one for the neutral state.}
\end{figure}

\begin{figure*}[ph!]\vspace{-1mm}
\begin{minipage}[h!]{0.5\textwidth}
\setlength{\tabcolsep}{8pt}
\captionof{table}{Ground State Energy differences}
\begin{tabular}{c|c|c|c}
$E^{1^+}-E^0$ & $E^{2^+}-E^0$ & $E^{1^+}_{\rm{rel}}-E^0$ & $E^{2^+}_{\rm{rel}}-E^0$ \\[1pt]
\hline
8.17 eV & 22.21 eV & 1.88 eV & 14.99 eV
\end{tabular}
\label{tab:tot_energy}
\end{minipage}\begin{minipage}[h!]{0.5\textwidth}
\setlength{\tabcolsep}{8pt}
\captionof{table}{Homo energies PBE and PBE+ADSIC}
\begin{tabular}{c|c}
$\epsilon_{\rm H}^{\rm{PBE}}$ & $\epsilon_{\rm H}^{\rm{PBE+ADSIC}}$ \\[1pt]
\hline
 -5.57 eV & -7.40 eV
\end{tabular}
\label{tab:homo_pbe}
\end{minipage}
\vspace{6mm}

\centering
\includegraphics[width=0.64\textwidth]{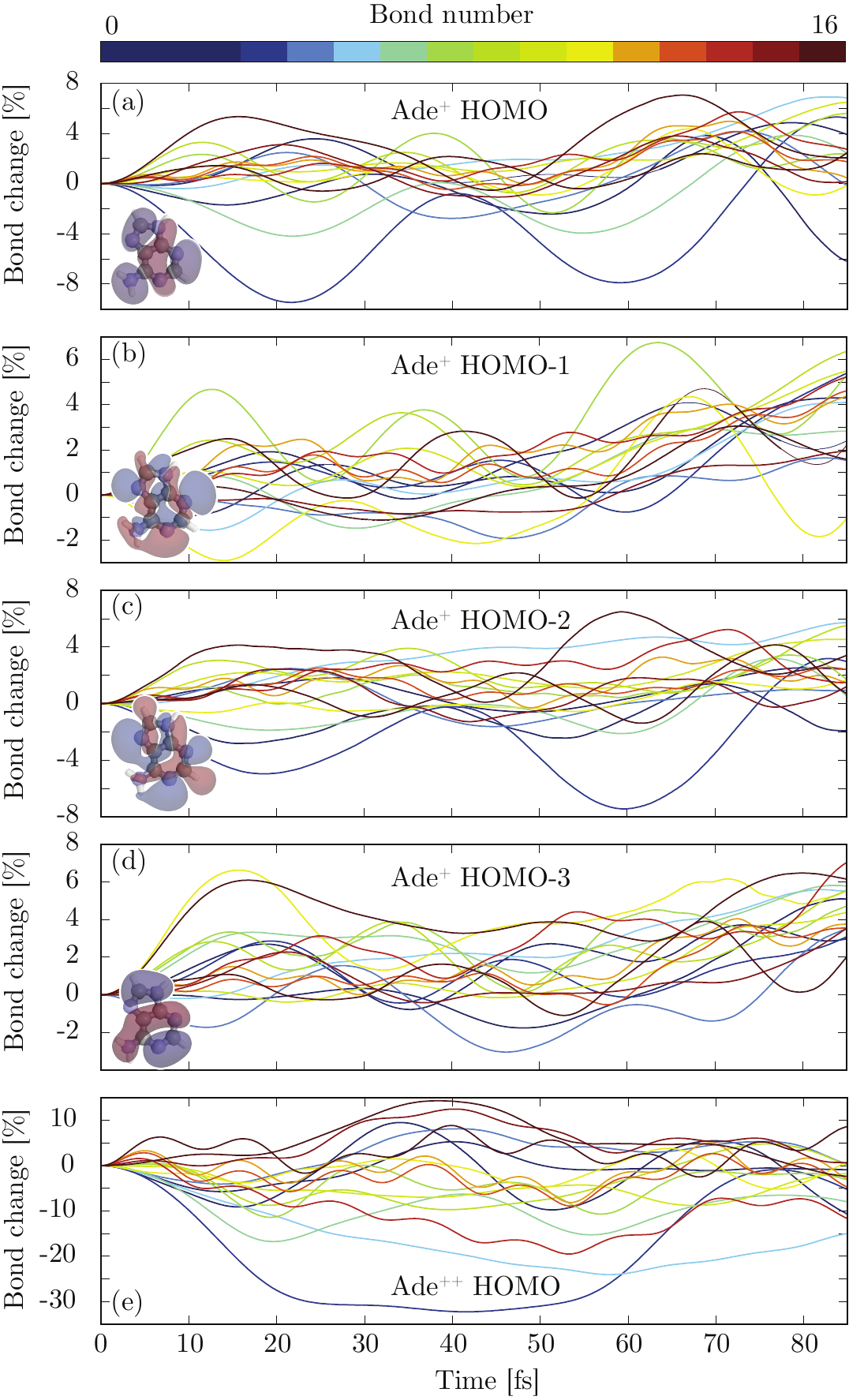}\vspace{-1mm}
\caption{{\bf Bond length variations following sudden Ionisation with TDDFT}. Adenine cations are time propagated following a sudden removal of an electron from different orbitals of the cation Ade$^+$, (a) HOMO, (b) HOMO-1, (c) HOMO-2, (d) HOMO-3 and two electrons from the HOMO (d). In all situations it appears that sudden ionisation leads to substantial bond elongation and potential fragmentation. In all the cases, negligible bond elongation is observed in the first 3 fs. Illustrations of the wavefunction from which the electron is removed are shown on the left side of panels (a)--(d).}
\label{fig:td_bonds}
\vspace{-1mm}
\end{figure*}

Ground state ab-initio characterisation of the Adenine molecule (sketched in Fig.~\ref{fig:AdCartoon}) has shown that if the geometry is allowed to relax, it is possible to find a stable molecule both for the singly and doubly ionised state. 
The effect of the geometry relaxation on the charged states of Adenine is relatively small as it can be seen from Fig.~\ref{fig:bonds_go_charged} for the singly (a) and doubly (b) ionised case. In such a figure, we report the relative variation (with respect to the optimised neutral structure) of the bond lengths at each optimisation step. 
To quantify the importance of geometry relaxation in the stabilisation of ionised adenine, we evaluated the total energy difference of singly (Ade$^+$) and doubly charged (Ade$^{2+}$) state with respect to the neutral (Ade) state. The values are reported in Tab.~\ref{tab:tot_energy} for the unrelaxed and relaxed structures.
We can infer that the minimum energy that the XUV+IR pulses have to transfer to the system in order to reach the stable dication state is $\sim 15$ eV, which is in agreement with the experimental observation that no stable dication is observed if the high energy part of the XUV spectrum is filtered out.

The calculations have been performed with the Octopus code\cite{Andrade2015} within a DFT framework where a PBE functional has been used. A grid spacing of $h=0.28$ Bohr has been used for the simulations.

To reproduce the first ionisation energy of adenine reported in the experimental literature of 8.48\;eV\cite{Lin:2002br} we applied the so called averaged density self-interaction correction (ADSIC)\cite{0953-4075-35-4-333} which corrects the Coulomb potential tails poorly described by the PBE functional.  As it can be noted from Tab.~\ref{tab:homo_pbe} the ionisation energy (which according to Koopman's theorem is equivalent to the energy of the HOMO) calculated with the SIC gets much closer to the experimental one compared to the uncorrected one. For the following results all the calculations are performed with the PBE-SIC functional.

%
%


\subsection{Bond elongation}

In this section we demonstrate that ionisation of adenine leads to substantial bond elongation and therefore possible dissociation. For such a proof we perform a sudden single and double ionisation and we let the electronic and nuclear system propagate in time by means of TDDFT+Ehrenfest dynamics\cite{Alonso:2008fz,Andrade:2009gaa}. In our calculations, the removal of electrons is orbital specific, in Fig.~\ref{fig:td_bonds}(a)--(d) we show the bond length dynamics for singly ionised adenine where the electron has been suddenly removed at the beginning of the time propagation from a given orbital.
In all the outer valence orbitals investigated the ionisation leads to substantial bond elongation. 
It is worth noting that in all the presented cases, bond elongation is negligible in the the first 3 fs after ionisation.
For completeness in Fig.~\ref{fig:td_bonds}(d) we also report the bond length dynamics in the case adenine is doubly ionised by removing 2 electrons from the HOMO. As expected, the bond rupture is faster in this case.

\begin{figure}[!t]
\includegraphics[width=1\columnwidth]{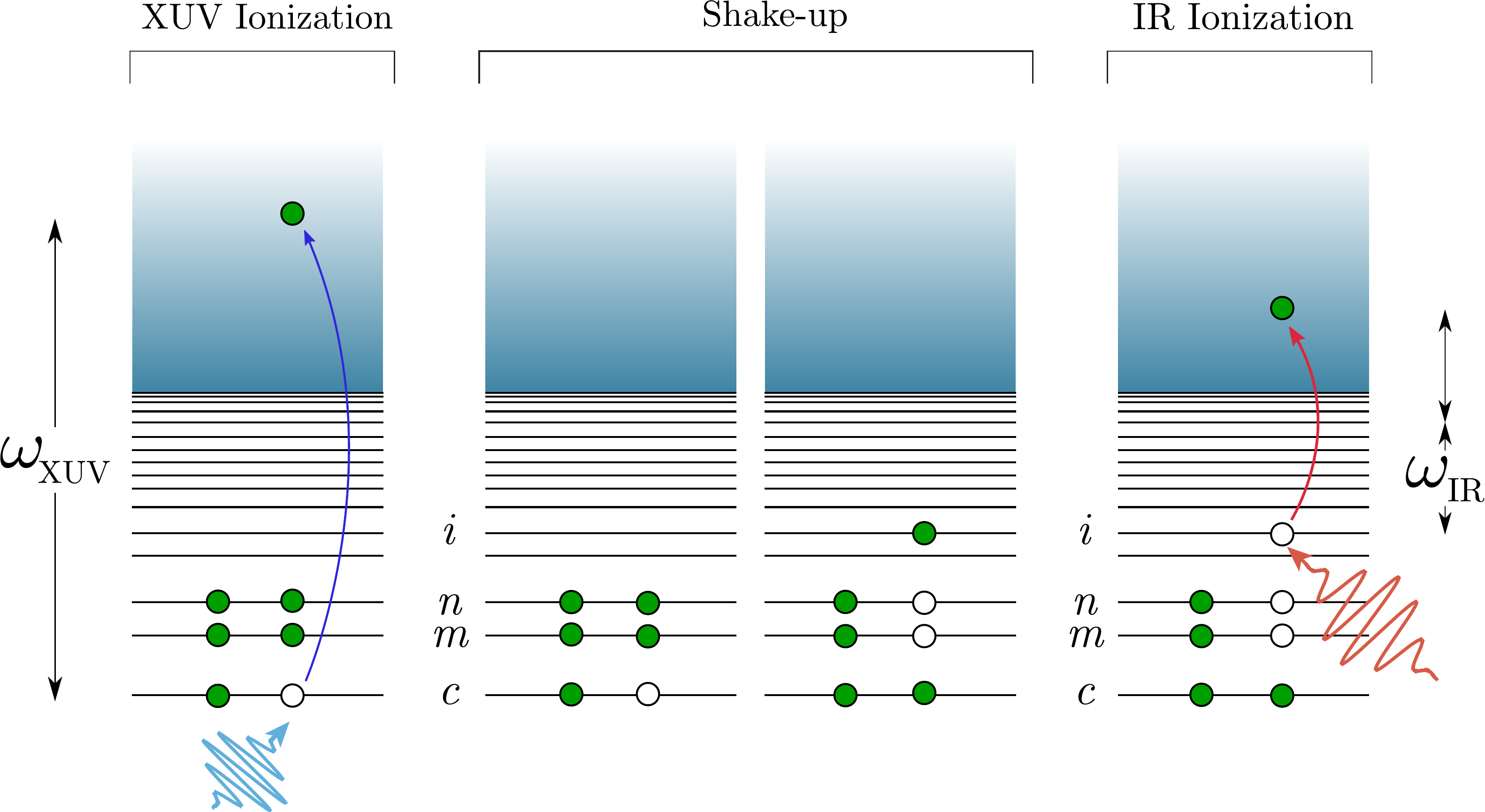}\vspace{-1mm}
\caption{\label{fig:shakeup}
\textbf{Three step model for the creation of stable adenine dication}. The molecule is ionised by the XUV pulse leaving it with hole in the core state $c$. During the shake-up process, this core-hole moves up to the valence region, through the excitation of an electron from the occupied level $n$ (or $m$) to the unoccupied level $i$ and the refilling of the core hole by an electron from level $m$ (or $n$). Finally the IR pulse ionises an electron from this excited level, leaving the molecule doubly ionised.}
\end{figure}

We conclude that an internal electronic relaxation via a correlated mechanism is needed for the adenine molecule to reach a stable double ionised form.

\subsection{Modelling the internal electronic relaxation process with rate equations}

\subsubsection{Shake-up matrix elements}

Owing to electron-electron correlation an ionised system can undergo an electronic rearrangement, in which previously unoccupied bound states become populated through a so called shake-up process, which is depicted in the central panel of Fig.~\ref{fig:shakeup}.

Driven by Coulomb interaction, the electron in $\phi_{m}$ fills the hole in $\phi_{c}$ and transfers the energy to the electron in $\phi_{n}$ which is then excited to the state $\phi_{i}$.
Overall, the shake-up process has to conserve energy, i.e. $\epsilon_i - \epsilon_n = \epsilon_m - \epsilon_c$. Here we are interested in calculating the rate for such a process for the adenine molecule following XUV photoionisation. Using Fermi's golden rule such a rate is given by \vspace{-1mm}\begin{equation}
W_{FI} = 2\pi |\langle \Psi_F| \hat{V}_C | \Psi_I\rangle |^2 \delta(\epsilon_F-\epsilon_I) .\label{eq:FGR}
\end{equation}\vspace{-1mm}
tried to place the following {widetext} equation differently
where $\Psi_I$ and $\Psi_F$ are the initial and final many-body wavefunctions, $\hat{V}_C$ the Coulomb interaction operator and the Dirac delta ensures energy conservation.
Let us first restrict our analysis to the four specific single particle states $\phi_c, \phi_m, \phi_n, \phi_i$. In such a basis, the Coulomb interaction, that we consider as the perturbation that drives the shake-up, can be written (in second quantization) as:\vspace{-1mm}
\begin{align}\nonumber
    \hat{V}^C_{mn} = \frac{1}{2}\sum_{\sigma\sigma'\in\{\uparrow\downarrow\}}  \Big\{  & v_{cimn}c^{\dagger}_{c\sigma}c^{\dagger}_{i\sigma'}c_{m\sigma'}c_{n\sigma} + \\[-2mm]
    &   v_{icmn}c^{\dagger}_{i\sigma}c^{\dagger}_{c\sigma'}c_{m\sigma'}c_{n\sigma} + h.c.\Big\},
\end{align}\vspace{-1mm}
where we have defined the Coulomb integrals as:
\begin{equation}
    v_{ijmn} = \int d{\bf{r}}d{\bf{r}}' \phi_i^*({\bf{r}})\phi_j^*({\bf{r}}')\frac{1}{|\bf{r}-\bf{r}'|}\phi_m({\bf{r}}')\phi_n({\bf{r}}).
    \label{coulomb-int}
\end{equation}

In our picture, the initial state of a shake-up process is the one resulting from the XUV photo-ionisation and, without loss of generality, we choose it to be $|\Psi_I\rangle = c_{n\uparrow}|\Psi_0\rangle$, with $|\Psi_0\rangle$ consisting of doubly occupied $i,m$ and $n$ orbitals. 
The application of the Coulomb perturbation on such an initial state can be evaluated as follows:
\begin{widetext}
\begin{equation}
    \begin{split}
        \hat{V}^C_{mn} | \Psi_I\rangle &= \frac{1}{2}\sum_{\sigma\sigma'\in\{\uparrow\downarrow\}}\left\{v_{cimn}c^{\dagger}_{c\sigma}c^{\dagger}_{i\sigma'}c_{m\sigma'}c_{n\sigma} + v_{icmn}c^{\dagger}_{i\sigma}c^{\dagger}_{c\sigma'}c_{m\sigma'}c_{n\sigma} + h.c.\right\} c_{n\uparrow}|\Psi_0\rangle \\
        &= \frac{1}{2}\sum_{\sigma\in\{\uparrow\downarrow\}}\left\{v_{cimn}c^{\dagger}_{i\sigma}c^{\dagger}_{c\uparrow}c_{n\uparrow}c_{m\sigma} + v_{icmn}c^{\dagger}_{i\sigma}c^{\dagger}_{c\uparrow}c_{m\uparrow}c_{n\sigma}\right\} c_{n\uparrow}|\Psi_0\rangle
    \end{split}
\end{equation}
\end{widetext}

where the h.c. does not contribute because it contains terms of the type $c^{\dagger}_{m\sigma}c^{\dagger}_{n\sigma'}$. Knowing the result of the Coulomb operator on the initial state and assuming the final state to be $|\Psi_F\rangle = |\Psi_i^{mn}\rangle = c^{\dagger}_{i\uparrow}c_{m\uparrow}c_{n\uparrow}|\Psi_0\rangle$ (again, without loss of generality) we can apply the Fermi golden rule in Eq.~(\ref{eq:FGR}) and write:\vspace{-2mm}
\begin{equation}
    R^{\rm{Sh-up}}_{imnc} = \frac{\pi}{2} ~ \left|v_{icmn} + v_{cimn}\right|^2 \delta(\epsilon_i - \epsilon_n - (\epsilon_m - \epsilon_c))\\[1mm]
\end{equation}

In general for a given initial core-hole state $c$ and a final occupied excited state $i$ there might be several $m$ and $n$ state combinations compatible with the shake-up process. For this reason the rate of the shake-up independent of the orbital localization of the two final holes is given by:\vspace{-2mm}
\begin{equation}\label{eq:su_me}
    R^{\rm{Sh-up}}_{ic} = \frac{\pi}{2} \!\sum_{n\ge m}\left|v_{icmn}+v_{cimn}\right|^2 \delta(\epsilon_i - \epsilon_n - (\epsilon_m - \epsilon_c)).\\[1mm]
\end{equation}
Note that we restricted the sum to $n\ge m$ to avoid the double counting that would arise from the fact that the holes are identical.
For practical purposes the delta function in the formula above is replaced by a Lorentzian, $L_{\epsilon_j}^{\gamma_j}(\omega) = \frac{1}{2\pi}\frac{\gamma_j}{(\omega-\epsilon_j)^2 + (\gamma_j/2)^2}$, as follows:\vspace{-1mm}
\begin{equation}
R^{\rm{Sh-up}}_{ic} \!=\! \frac{\pi}{2} \! \sum_{n\ge m}
  \left|v_{icmn}+v_{cimn}\right|^2
  L_{\epsilon_m + \epsilon_n-\epsilon_i-\epsilon_c}^{\eta}(\omega\!=\!0),
\end{equation}
where the width $\eta$ enters as a parameter.

\subsubsection{Ionisation rate: evaluating orbital dependent cross-sections}
The initial state in the shake-up process is created by the XUV pulse (see leftmost panel of Fig.~\ref{fig:shakeup}) and the probability of creating the initial hole in a state $\phi_i$ is determined by the XUV photoionisation probability of such a state which is given by:
\begin{equation}
    \begin{split}
        &P^{\rm{Ion}}_{i}(\omega) =2\pi n_{\rm{spin}} \sum_{\bf{k}}\left|\langle \phi_{\bf{k}} | e^{i{\bf q} \cdot {\hat{r}}} | \phi_i \rangle\right|^2 \delta(\omega - (\epsilon_{\bf{k}}-\epsilon_i)) \\
        &=\frac{2\pi}{V}\sum_{\bf{k}} \left|\int d{\bf{r}} e^{i(\bf{k+q})\cdot\bf{r}}\phi_i(\bf{r})\right|^2\delta\left(\frac{k^2}{2}-(\omega + \epsilon_i)\right),
    \end{split}
\end{equation}
where $\omega$ is the energy of the perturbation, $|\phi_{\bf{k}}\rangle$ the outgoing electron wavefunction, which in the second step we assume to be a planewave, and $A_Ie^{i{\bf q} \cdot {\hat{r}}}$ the spatial component of the electromagnetic field. 
In the long wavelength limit (${\bf{q}}\rightarrow 0$) we can recognize the Fourier transform of the wavefunction $\tilde{\phi}_i({\bf{k}}) = \int d{\bf{r}} e^{i(\bf{k})\cdot\bf{r}}\phi_i(\bf{r})$ and write:\vspace{-1mm}
\begin{equation}
    \label{eq:orbital_IR}
    P^{\rm{Ion}}_{i}(\omega) = \frac{2\pi}{V}\sum_{\bf{k}} \left|\tilde{\phi}_{i{\bf k}}\right|^2 \delta\left(\frac{k^2}{2}-(\omega + \epsilon_i)\right).
\end{equation}\vspace{-1mm}

\begin{figure}[b!]
\label{fig:ionization_rate}
\centering
\includegraphics[width=0.65\columnwidth]{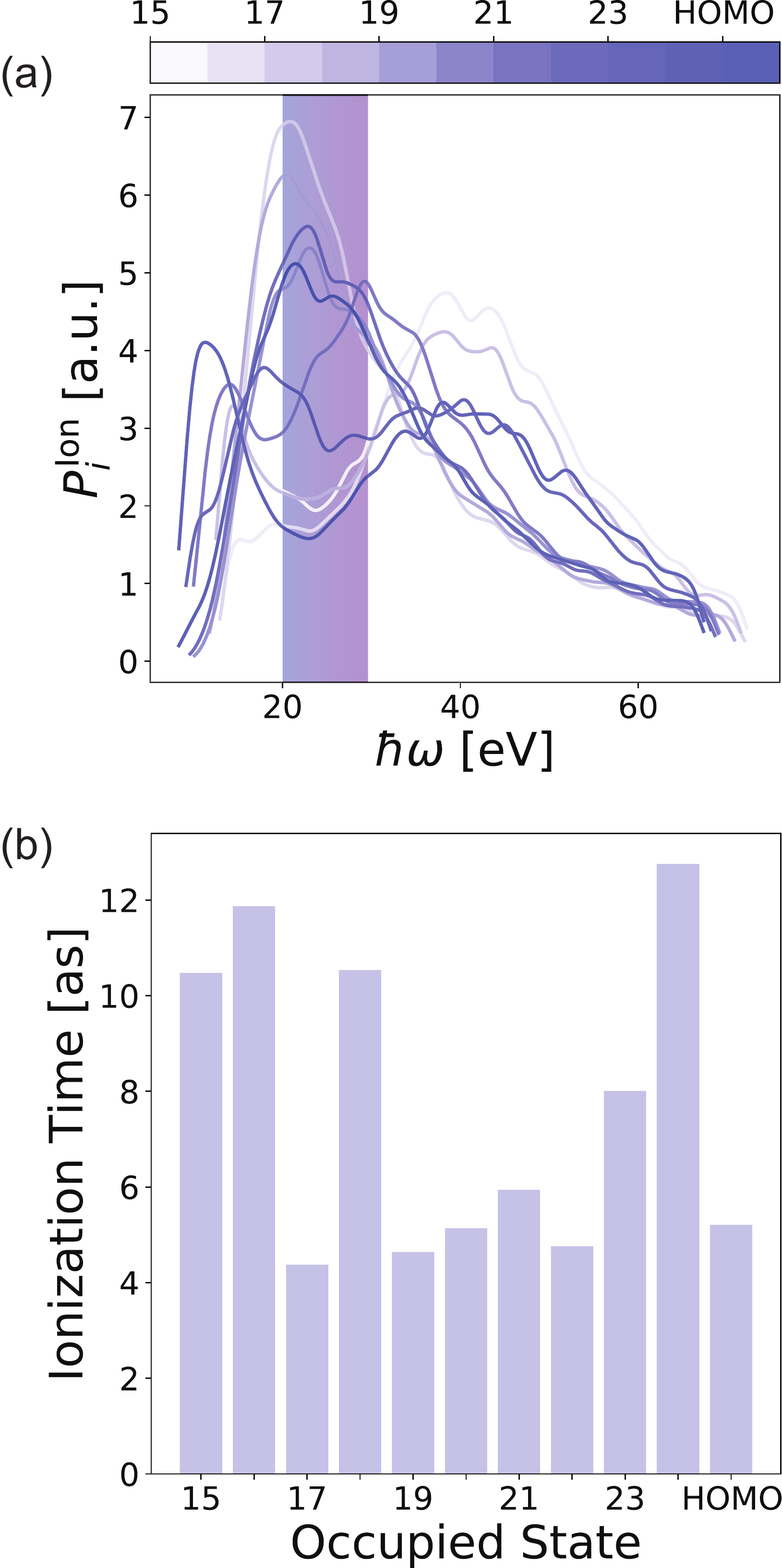}
\caption{\textbf{XUV-photoionisation rate for the occupied orbitals} (a) calculated with Eq.~(\ref{eq:orbital_IR}). (b) Orbital dependent ionisation time calculated by integrating the curves in (a) within the XUV energy window.}
\end{figure}

Using this formula, we evaluated the orbital dependent ionisation probability for the 15 outer valence KS orbitals reported in Fig.~\ref{fig:ionization_rate}(a).

By convoluting these photoionisation probabilities with an energy distribution function $f^{\rm{XUV}}(\omega)$ for the XUV pulse, we can get the orbital resolved ionisation rate as:
\begin{equation}
    R_c^{\rm{XUV}} = \int d\omega \, f^{\rm{XUV}}(\omega) \, P_c^{\rm{Ion}}(\omega).
\end{equation}
In our calculations we make the easiest assumption by choosing $f^{\rm{XUV}}(\omega)$ as a stepwise window function between $19.5$ and $30.5$ eV (as in the experiment) and extract the time it takes for a given valence orbital to be fully ionised as the inverse of the ionisation rates, see Fig.~\ref{fig:ionization_rate}(b).

\subsubsection{The shake-up process and the characteristic time delay}

\begin{figure}[b]
\label{fig:nc_me_Ri}
\centering
\includegraphics[width=0.85\columnwidth]{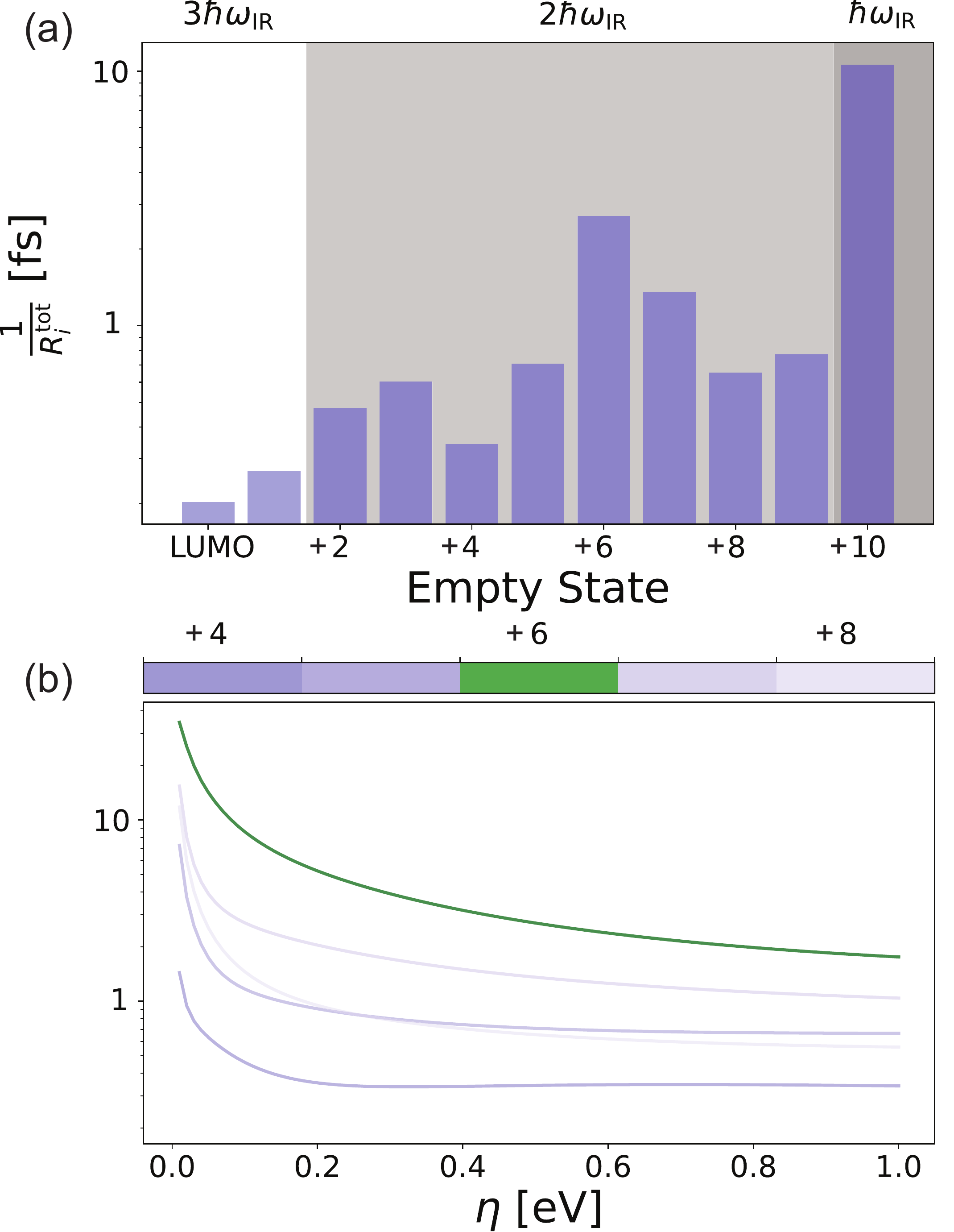}
\caption{\textbf{Shake-up rates for the first unoccupied orbitals:} (a) Shake-up rates where the Dirac delta has been approximated by a Lorentzian with width $\eta=0.1$ eV  and a photoionisation time ($T_{\rm{XUV}}=100$ as). (b) Same rates as a function of the broadening parameter $\eta$.}
\end{figure}

We can finally combine the XUV photoionisation probability with the state resolved shake-up rates and get the characteristic shake-up times. More specifically the probability that an empty bound state gets populated by shake-up processes following XUV ionisation can be written as:
\begin{equation}
P_{i}^{\rm{Sh-up}}(t) = 1 - e^{R^{\rm{tot}}_it}   
\end{equation}
where the total rate is given by:
\begin{equation}
    R^{\rm{tot}}_i = \sum_c \left(1-e^{-R_c^{\rm{XUV}}T^{\rm{XUV}}}\right)R_{ic}^{\rm{Sh-up}}
\end{equation}
i.e. the probability of having a hole in state $c$ times the rate of shake-up towards a state $i$ summed over all initial holes. The probability of creating initial holes takes into account the actual duration of the XUV pulse $T^{\rm{XUV}}$.

In Fig.~\ref{fig:nc_me_Ri}(a) we show the characteristic electronic shake-up times as reported in the main text. For more thorough validation of this result we computed the same quantities as a function of the artificial parameter $\eta$. These results are reported in Fig.~\ref{fig:nc_me_Ri}(b) for the final empty states ionised with two IR photons and it is apparent that the variation of the characteristic times with respect to $\eta$ is small for any $\eta>0.1$ eV which is safe to assume considering typical quasiparticle lifetimes in molecules \cite{Caruso2013}.

\FloatBarrier

\subsection{Non-equilibrium Green's function approach}
\subsubsection{Theoretical description}
In this section we discuss the details of the ab-initio real-time
propagation presented in the main paper. We use the non-equilibrium Green's 
function (NEGF) formalism and solve numerically the Kadanoff-Baym equations
(KBE's)~\cite{kadanoff1962quantum,svl-book,balzer2012nonequilibrium}
within the Generalized Kadanoff-Baym Ansatz 
(GKBA)~\cite{PhysRevB.34.6933}.

\begin{figure}[t]
  \centering
  \includegraphics[clip, width=0.8\columnwidth]{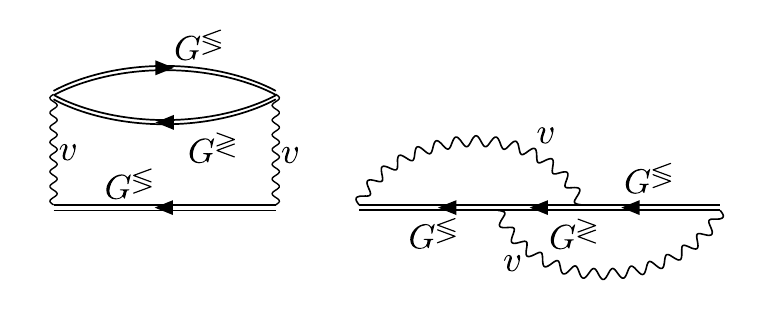}\vspace{-1mm}
  \caption{{\bf Diagrammatic representation of the second Born
  self-energy}. The double line represents the Green's 
  function $G^\lessgtr$ while the wiggly line represents
  the Coulomb interaction $v$. }
  \label{fig:diagram}
\end{figure}

Within the GKBA the KBE's collapse into a single equation for 
the single-particle density matrix $\rho(t)$
\begin{equation}\label{eqtosolve}
     \frac{d}{dt}\rho(t) + i[h_{\rm HF}(t), \rho(t)] =  - 
     [I_{\mathrm{corr}}(t) - I_{\mathrm{ion}}(t) + \mathrm{h.c.}].
\end{equation}
The left-hand side contains the commutator between $\rho$ and the time-dependent 
Hartree-Fock (HF) Hamiltonian 
\be
h_{\rm 
HF}(t)=h_{\mathrm{KS}}-V_{\mathrm{Hxc}}+V_{\mathrm{HF}}(t)+\bcallE(t) 
\cdot \bld  \, ,
\label{hfham}
\ee
where $h_{\mathrm{KS}}$ is the Kohn-Sham (KS) Hamiltonian, 
$V_{\mathrm{Hxc}}$ is the Hatree-exchange-correlation potential (both 
$h_{\mathrm{KS}}$ and $V_{\mathrm{Hxc}}$ are evaluated at 
the equilibrium KS density),
$V_{\mathrm{HF}}(t)=V_{\mathrm{HF}}[\rho(t)]$ is the time-dependent HF 
potential,
$\bcallE(t)$ is the external pulse, and $\bld=(d_{x},d_{y},d_{z})$ is the molecular dipole 
vector.
Correlation effects are included in the right-hand side of 
Eq.~(\ref{eqtosolve}) via the {\it 
collision integral}
\begin{align}
    I_{\mathrm{corr}}(t) = \int_0^t dt'  [&\Sigma_{\mathrm{corr}}^> (t,t') G^<(t',t)\nonumber\\
    - &\Sigma^<_{\mathrm{corr}}(t,t') G^>(t',t)] \, ,
\end{align}
where $G^{\lessgtr}$ is the lesser (greater) non-equilibrium Green's 
function,
and  $\S_{\mathrm{corr}}^{\lessgtr}$ is the correlation lesser (greater) non-equilibrium
self-energy. Here we approximate $\S_{\mathrm{corr}}$  at the second Born (2B) level, 
see Fig.~\ref{fig:diagram} for its diagrammatic representation.
The GKBA consists in approximating the lesser and greater Green's 
functions  as
 \begin{align}
     G^<(t,t') & = - G^R(t,t')\rho(t') + \rho(t) G^A(t,t'),\\
     G^>(t,t') & = G^R(t,t')\bar{\rho}(t') - \bar{\rho}(t) G^A(t,t'),
 \end{align}
 where $\bar{\rho}(t)=1- \rho(t)$ and
 \begin{equation}
     G^R(t,t')=[G^A(t',t)]^\dag= -i \theta(t-t')
     T[e^{-i\int_{t'}^t d \bar{t}\, h_{\rm HF}(\bar{t})}].
 \end{equation}
The second term appearing on the right-hand side of Eq.~(\ref{eqtosolve}) is the 
{\it ionisation integral} 
\begin{equation}
    I_{\mathrm{ion}}(t) = \int_0^t dt'  \Sigma_{\mathrm{ion}}^> 
    (t,t') G^<(t',t) \, ,
\end{equation}
with $\Sigma_{\mathrm{ion}}^>$ the ionisation self-energy. This term describes
the excitation of bound electrons to continuum states occurring during
the ionisation process induced by the pulse $\bcallE(t)$.

The numerical solution of  
Eq.~(\ref{eqtosolve}) with $\S_{\rm corr}$ in the 2B 
approximation scales quadratically 
with the maximum propagation time.
This favourable scaling (in comparison to the cubic scaling of the KBE) 
allows us to follow the 
dynamics of molecules with $\sim 10^{2}$  active electrons for tens of 
femtoseconds.

\subsubsection{Computational method}

In this section we discuss how Eq.~(\ref{eqtosolve}) is practically solved
to study the ultrafast electron dynamics of adenine after the action of a XUV pulse.
We first obtain the KS ground state (GS) of the molecule using the
Octopus code \cite{Andrade2015}, with HSCV pseudopotentials \cite{PhysRevB.32.8412},
the PBE approximation for the exchange-correlation potential
\cite{PhysRevLett.77.3865} and the averaged density self-interaction
correction (ADSIC) \cite{0953-4075-35-4-333}. The grid is a sphere 
of radius $20$ Bohr with spacing $0.25$ Bohr.
The lowest 57 KS orbitals are bound and constitute the {\it 
active space} to describe the molecular $\rho$. The remaining KS 
orbitals (with KS energy $\e^{\rm KS}_{k}>0$) up to energies $\sim 41$\;eV (about 3000 states) are 
instead used to build $\S_{\mathrm{ion}}$.

\begin{figure}[b]
\centering\vspace{-1mm}
\includegraphics[clip, width=1\columnwidth]{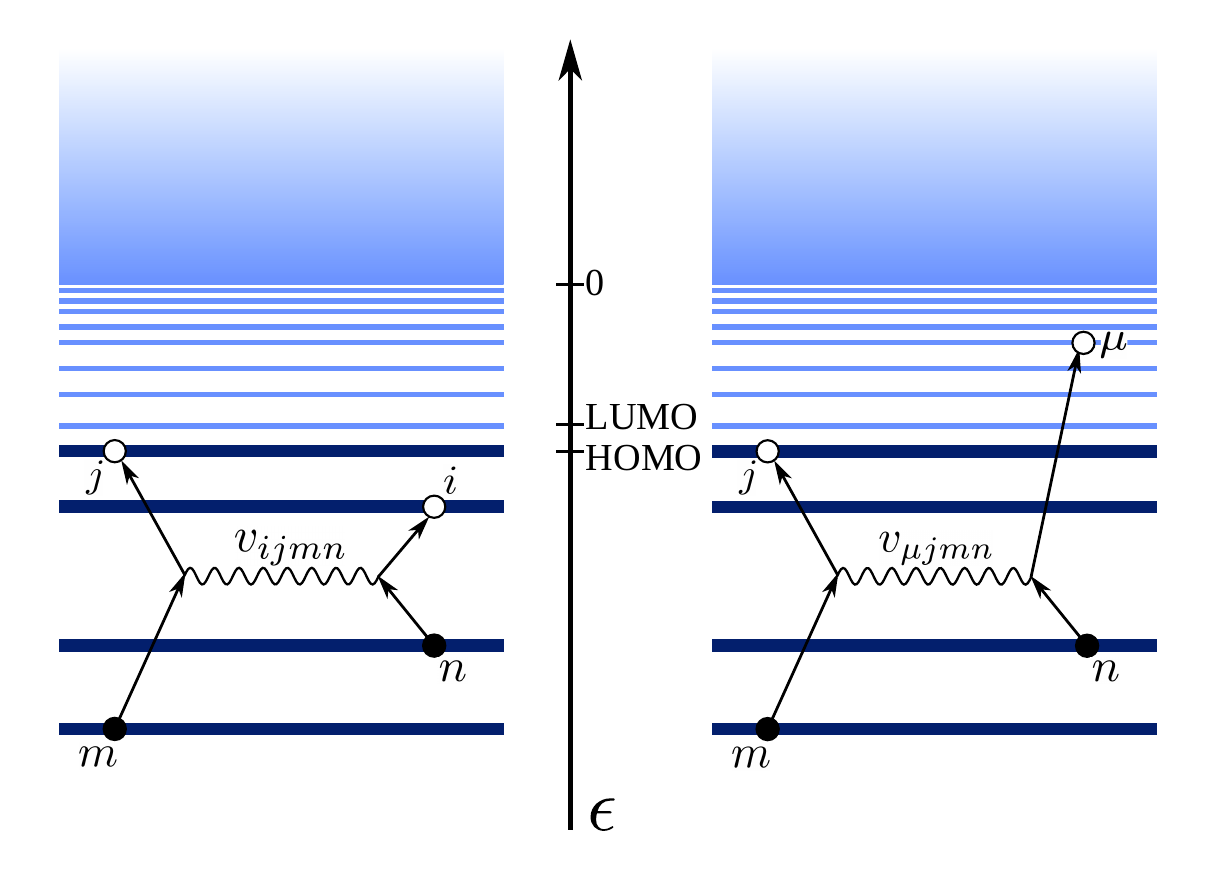}\vspace{-2mm}
\caption{{\bf Schematic illustration of scattering processes}.
Left: scattering involving electrons in initially occupied states (labelled by roman indices).
Right: shake-up scatterings leading to the transition of an electron in an   initially unoccupied state (labeled by $\m$). The initially occupied and unoccupied states are represented by a dark blue and a light blue line, respectively. The shaded area represents the continuum -- no scattering is considered in this area.}
\label{fig:scatterings}
\end{figure}

We then solve the equilibrium HF problem (with $\bcallE=0$) in the 
active space. Since correlation effects are typically weak in the ground state 
of biomolecules~\cite{Kuleff2005,KULEFF2007320,Ruberti2018}, we set $\rho(0)=\rho_{\mathrm{HF}}$
as initial condition for Eq.~(\ref{eqtosolve}).
It is therefore convenient to work in the HF basis 
$\{\vf^{\mathrm{HF}}_{i}(\blr)\}$. In this basis the equilibrium HF 
Hamiltonian $h_{\rm HF}$ is 
diagonal and the 2B self-energy reads
\begin{align}\label{2B}
     \Sigma^{\lessgtr}_{\mathrm{corr,}ij}(t,t')=
      \sum_{nmpqsr}& v_{irpn}
    (2v_{mqsj}-v_{mqjs})\\ 
    & \times G^{\lessgtr}_{nm}(t,t')G^{\lessgtr}_{pq}(t,t')
     G^{\lessgtr}_{sr}(t',t),\nonumber
\end{align}
where the Coulomb integrals $v_{irpn}$ are computed like in \eqref{coulomb-int} with the HF basis functions.
The correlation self-energy in Eq.~(\ref{2B})
is numerically evaluated using the {\it dissection 
algorithm}~\cite{PS.2018} implemented in the CHEERS code~\cite{Perfetto2018}.
Additionally
only those Coulomb integrals with at most one index 
above the HF-HOMO are retained, see Fig.~\ref{fig:scatterings}. This amounts to 
including only shake-up processes for electrons visiting  
HF states initially unoccupied. A similar approximation has been 
recently used in the context of Auger decays~\cite{Covito2018},
finding excellent agreement with Configuration Interaction 
calculations.
The approximation reduces the number of 
Coulomb integrals involved in $\S_{\rm corr}$: 
from $N^4_{\rm tot}$, where $N_{\rm tot}=57$ is
the dimension of the active space, to $N^4_{\rm occ} + 4 N^3_{\rm occ}
(N_{\rm tot} - N_{\rm occ})$, where $N_{\rm occ}=27$ is the number of
initially occupied states; hence about an order of magnitude less.
Furthermore, $\rho(0)=\rho_{\mathrm{HF}}$ is a stationary solution of 
Eq.~(\ref{eqtosolve}) in the absence of external fields.

The ionisation self-energy in the HF basis reads
\begin{equation}\label{sigmaion}
  \Sigma^>_{\mathrm{ion},ij}(t,t')
  = -i \hspace{-2ex} \sum_{k:~\ve^{\mathrm{KS}}_{k}>0} \hspace{-2ex}
    [\bcallE(t)\cdot
      \bld_{ik}]e^{-i 
      \epsilon^{\mathrm{KS}}_{k}(t-t')}[\bcallE(t')\cdot \bld_{kj}],
\end{equation}
where
\be
 \bld_{ij}= \int d\blr ~ \vf^{\mathrm{HF}}_{i}(\blr) \blr  \vf^{\mathrm{HF}}_{j}(\blr),
\ee
and $\ve^{\mathrm{KS}}_{k}=\ve^{\mathrm{HF}}_{k}$ since we discard 
the Coulomb repulsion of HF electrons in the continuum. The main 
physical process left out by this further approximation is the Auger 
decay which, however, can be safely ignored for XUV pulses. We 
also emphasise that the ionisation self-energy in Eq.~(\ref{sigmaion})
is suitable only for  single-photon ionisation 
processes (like the ones induced by the XUV pulse of the 
present experiment). The inclusion of multiphoton processes would 
lead to a different dependence of  $\S_{\rm ion}$ on $\bcallE$.

The NEGF-GKBA Eq.~(\ref{eqtosolve}) is solved numerically using the 
CHEERS code~\cite{Perfetto2018}. From the single-particle density matrix 
we can easily calculate the time-dependent 
electronic density according to
\be
n_{\mathrm{el}}(\blr, t)=\sum_{ij}  \vf^{\mathrm{HF}}_{i}(\blr) \rho_{ij}(t)  
\vf^{\mathrm{HF}}_{i} (\blr) .
\ee

\subsubsection{Ionisation}
\begin{figure}[b!]
\centering
\includegraphics[clip, trim=10mm 10mm 5mm 1mm, width=1\columnwidth]{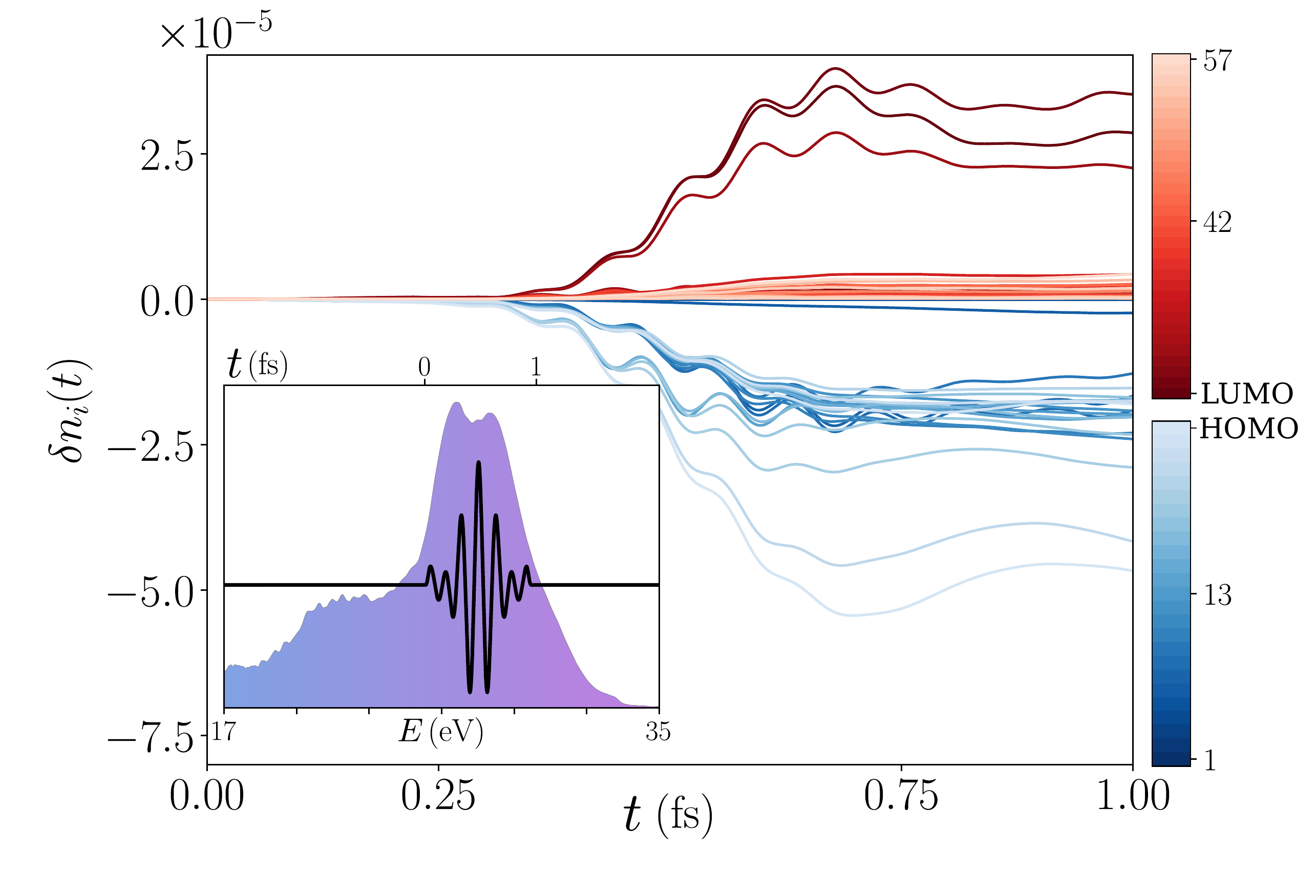}\vspace{-1mm}
\caption{ {\bf Time-dependent variation of the population}. Change of the population of the occupied (blue) and unoccupied (red) bound KS states of adenine excited by the XUV pulse perpendicular to the plane of the molecule.
The inset shows the spectrum and the temporal profile of the pump pulse.}
\label{fig:pumpaction}
\end{figure}

In our simulations we have studied the electron dynamics of adenine under the influence of the experimental XUV laser pulse. The
latter has a full-width at half-maximum of $\sim300~{\rm as}$ and  a
central frequency of $\sim27~{\rm eV}$ (see the inset of Fig.~\ref{fig:pumpaction}
for the temporal shape of the pulse and for its power spectrum). 
In Fig.~\ref{fig:pumpaction}
we show the time-dependent variation of the occupations $\d 
n_{i}(t)=n_{i}(t)-n_{i}(0)$ of the
bound KS orbitals $\{ \vf^{\mathrm{KS}}_{i} \}$  during 
the ionisation process.
The occupations  $n_{i}(t)$ are  obtained according to
\be
n_{i}(t)=\sum_{mn} U_{im} \rho_{mn}(t) U^{\dag}_{ni},
\ee
where  $U_{im}=\int d\blr~  \vf^{\mathrm{KS}}_{i}(\blr) 
\vf^{\mathrm{HF}}_{m}(\blr)$ is the change of basis transformation 
matrix.

The inspection of Fig.~\ref{fig:pumpaction} shows that  all 
initially occupied (unoccupied) KS states depopulate (populate) 
with superimposed oscillations following the cycles of the laser 
pulse. We point out that the initially
unoccupied KS states get populated mainly through  ultrafast 
shake-up processes. Longer timescale dynamics are discussed
in the main paper. 

\subsubsection{Charge migration}

\begin{figure}[b!]
\centering
\includegraphics[clip, trim=10mm 10mm 10mm 10mm, width=1\columnwidth]{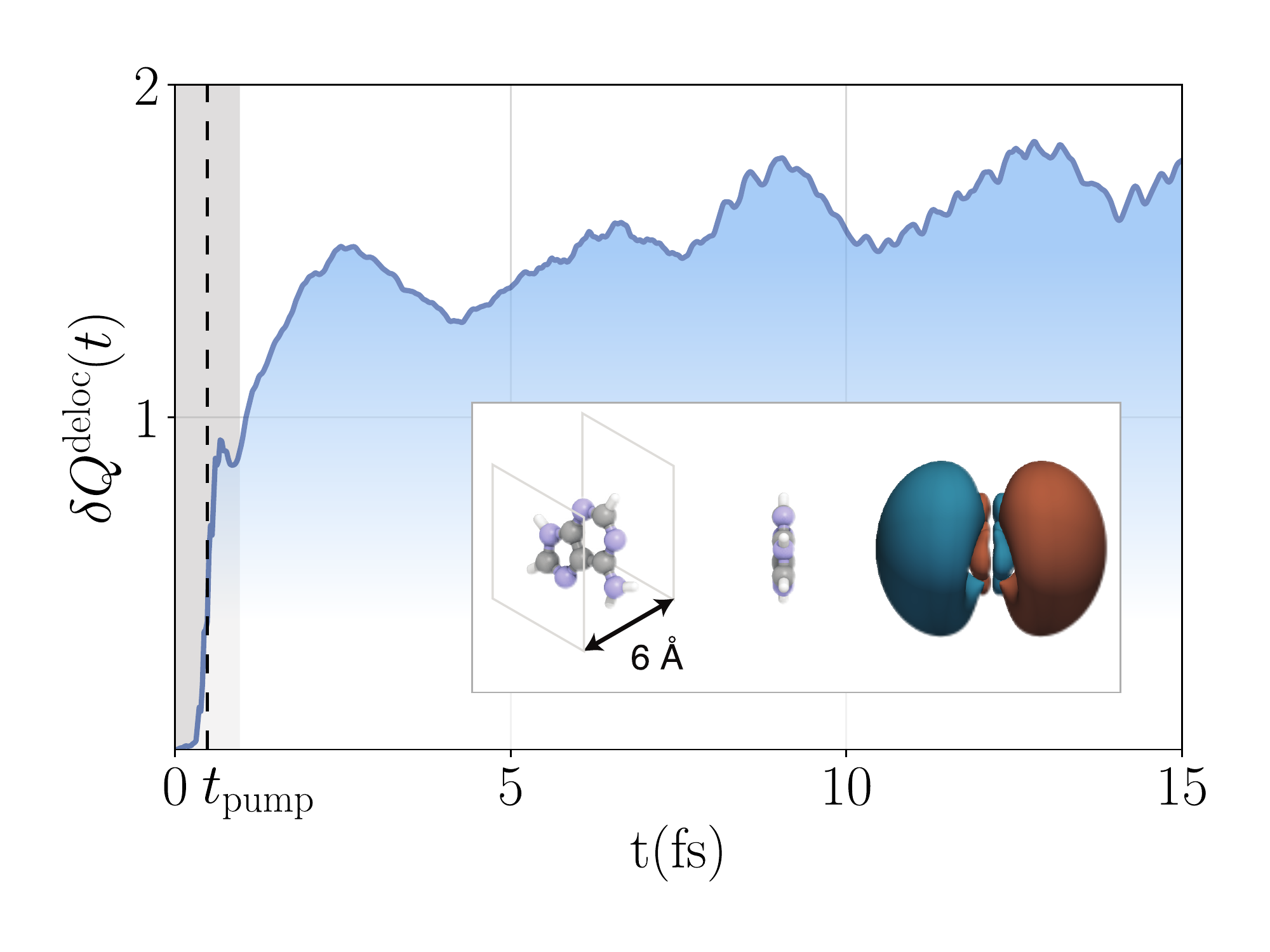}\vspace{-1mm}
\caption{{\bf Inflation.} Integrated time-dependent electron density more than $3~\AA$ away from the molecular plane. The grey shaded area represents the time-window of the pump pulse, having its peak at $t_{\rm pump}=0.48$\;fs. The y-axis has been rescaled by a factor of $10^5$. The inset shows the molecule and the planes defining the integration region (left) and the wavefunction of the special Kohn-Sham state LUMO+6 (right).}
\label{fig:integral_out}
\end{figure}

As stated in the main text, the simulations reveal an inflation of the electron density of the molecule over time, caused by the population of the particular KS state LUMO+6 on a timescale compatible with the delay observed in the yield of stable adenine dication. The inflation indicates that the electrons move away from the molecule over time. Defining a slab around the molecular plane of thickness $d = 6~\AA$ and integrating the electron density outside of this area, we have a measure of the molecular inflation. In Fig.~\ref{fig:integral_out} we show how this integrated electron density reflects the timescale of the inflation, as expected. The observed behaviour is consistent for different slab thicknesses.  This effect is due to the population of the delocalised bound states, as reported in the main text, and in fact has the two characteristic time scales observed in the population of the initially unoccupied KS states, the slower one being the one associated with the delayed shake-up transition.

\subsection{NIR-induced ionisation}

\begin{figure}[b]
\centering
\includegraphics[clip, trim=10mm 10mm 10mm 5mm,width=1\columnwidth]{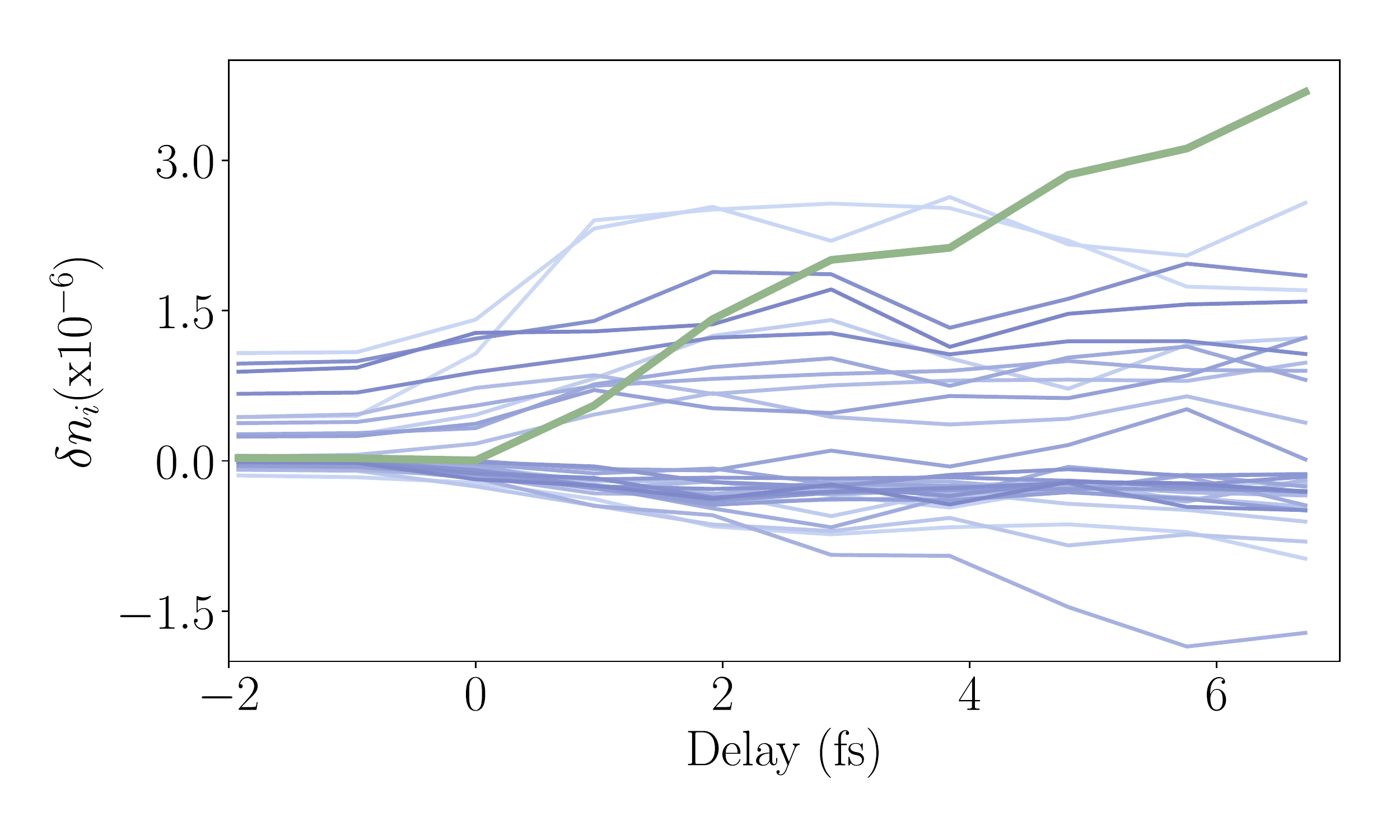}
\caption{{\bf Shake-up state depletion.} Average shake-up states depletion as a function of XUV-NIR delay. The average has been performed in a window of 1\;fs after the end of the pulses. In accordance to the main text, the LUMO+6 is drawn in green while the rest in shades of light blue.}
\label{fig:NIRionyield}
\end{figure}

To support the analysis on the role of the LUMO+6 state in the stabilisation dynamics of adenine we observe the behaviour of the system when the additional NIR pulse is introduced. In our simulation we have used a pulse with carrier frequency $1.7$\;eV and sin$^2$ envelope of total duration $200 \au \simeq 4.8$\;fs. Of particular interest is the time-dependent occupation of the shake-up states as the NIR pulse acts on the system. In fig.~\ref{fig:NIRionyield} the total NIR-induced state depletion for the different shake-up states as a function of XUV-NIR delay is shown. These are defined as the averaged value of the NIR-induced variation of the state occupation in a window of 1\;fs after the end of the NIR pulse. As stated in the main paper the NIR-induced depletion of the LUMO+6 state has an onset of 2--4\;fs and reproduces the behaviours of the adenine dication yield. Moreover, the LUMO+6 state is the only one, of all the shake-up states, showing these characteristics.

\FloatBarrier

\spacing{1.05} 
\let\oldbibliography\thebibliography
\renewcommand{\thebibliography}[1]{%
  \oldbibliography{#1}%
  \setlength{\itemsep}{0.9ex}%
}
